\documentclass[12pt,preprint]{emulateapj}

\slugcomment{accepted by ApJ}

\shorttitle{Accretion in quiescent 
SMBHs: optical study and interpretation}
\shortauthors{Soria et al.}

\begin{document}


\title{Accretion and nuclear activity of quiescent supermassive black
holes. II: optical study and interpretation}


\author{R. Soria\altaffilmark{1}, Alister W. Graham\altaffilmark{2},
	G. Fabbiano\altaffilmark{1},
	A. Baldi\altaffilmark{1}, 
	M. Elvis\altaffilmark{1},
	H. Jerjen\altaffilmark{2}, S. Pellegrini\altaffilmark{3},
	A. Siemiginowska\altaffilmark{1}}
\altaffiltext{1}{Harvard-Smithsonian Center for Astrophysics, 
	60 Garden st, Cambridge, MA 02138, USA}
\altaffiltext{2}{RSAA, Australian National University, Cotter Rd, ACT 2611}
\altaffiltext{3}{Astronomy Department, Bologna University, Italy}




\begin{abstract}
Our X-ray study of the nuclear activity in a new sample 
of six quiescent early-type galaxies, and 
in a larger sample from the literature, confirmed  
(Soria et al., Paper I) 
that the Bondi accretion rate of diffuse hot gas is not 
a good indicator of the supermassive black hole 
(SMBH) X-ray luminosity; in fact, the two 
quantities appear uncorrelated.
Here we suggest that a more reliable 
estimate of the accretion rate must include 
the gas released by the stellar population 
inside the sphere of influence of the SMBH 
(generally too small to be probed by {\it Chandra}), 
in addition to the Bondi inflow of hot gas 
across that surface.  We use optical surface-brightness 
profiles, from archival {\it  HST} images, to estimate 
the mass-loss rate from stellar winds 
in the nuclear region: we show that for our sample of 
galaxies it is an order of magnitude higher (typical values of 
$\sim 10^{-4}$--$10^{-3} M_{\odot}$ yr$^{-1}$) 
than the Bondi inflow rate of hot gas, as estimated 
from X-ray observations (Paper I).
Only by taking into account both sources of fuel 
can we constrain the true accretion rate, 
the accretion efficiency, and the power budget.
Radiatively efficient accretion is ruled out, 
for quiescent SMBHs. 
For typical radiatively inefficient flows, the 
observed X-ray luminosities of the SMBHs imply 
accretion fractions $\sim 1$--$10\%$ (i.e., $\sim 90$--$99\%$ 
of the available gas does not reach the SMBH) 
for at least five of our six target galaxies, 
and most of the other galaxies with known 
SMBH masses.
We discuss the conditions for mass conservation inside 
the sphere of influence of a SMBH, 
so that the total gas injection is balanced by accretion 
across the event horizon plus outflows.
We show that a fraction of the total accretion power 
(mechanical plus radiative) would be sufficient
to sustain a self-regulating, slow outflow
which removes from the nuclear region all the gas that does 
not sink into the BH (``BH feedback'').  
The rest of the accretion power may be carried out 
in a jet, or advected, depending on the details of the
radiative-inefficient accretion solutions. 
We also discuss possible scenarios 
that would lead to an intermittent nuclear  
activity, such as transitions in the outflow rate  
or in the efficiency.
\end{abstract}





\section{Introduction}

Most of the galaxies in the nearby universe 
have inactive nuclei (X-ray luminosities 
$\la 10^{40}$ erg s$^{-1}$). This may be due 
to an interplay of different factors: 
low rate of gas injection/inflow inside 
the ``sphere of influence''\footnote{Defined here 
as the region inside the Bondi accretion radius 
$r_{\rm acc} = 2GM_{\rm BH}/c^2_{\rm s}$, 
where the gravitational energy due to the SMBH dominates 
over the thermal energy of the gas; see also Section 4.3 
in Paper I.}  
of the supermassive black hole (SMBH); 
low fraction of the available gas 
being accreted onto the SMBH; low radiative efficiency 
of accretion, with the rest of the accretion power 
being advected, or carried out as mechanical 
luminosity by a jet or an outflow.

Our goal is to estimate these factors quantitatively, 
to discriminate between different radiatively-inefficient 
scenarios, and to outline the power and mass budget 
inside the sphere of influence.
To do so, we have selected a sample 
of six quiescent early-type galaxies for which
a direct, kinematical measurement of their SMBH
mass has been made (Table 1). Using new {\it Chandra} data, 
we have determined the X-ray luminosity 
of the nuclear sources, the density and temperature of the 
surrounding hot interstellar medium (ISM), and 
the classical Bondi rate ($\dot{M}_{\rm B}$) of inflow  
of the hot ISM into the SMBH sphere of influence. 
The main results are summarized in Table 1, see 
Soria et al.~(2005, hereafter Paper I) 
for details. We found that the nuclear sources are 
much fainter than predicted by the standard-disk scenario 
(which is also ruled out by other theoretical 
considerations at such low accretion rates).
However, they are brighter than predicted 
by radiatively-inefficient models, perhaps 
suggesting that the Bondi rate is underestimating 
the true accretion rate $\dot{M}$ (Table 1, Col.9).

We then considered another eighteen galaxies for which
the SMBH X-ray luminosity and the hot-ISM Bondi inflow rate 
have been calculated or constrained from previous work (Pellegrini~2005, 
and references therein; Garcia et al.~2005; David et al.~2005).
For most of these galaxies, the SMBH X-ray luminosity is lower than 
predicted by the standard ADAF model, suggesting 
that the true accretion rate $\dot{M} \ll \dot{M}_{\rm B}$.
Overall, there is little or no correlation 
between Bondi rate and X-ray luminosity of the SMBH 
(Fig.~14 in Paper I).

However, the hot, X-ray emitting ISM may represent  
only a small fraction of the gas fuelling the SMBH.
That can be the case in systems where gas can cool efficiently 
(cooling timescale $<$ accretion timescale), 
or, vice versa, if gas is injected into the inner regions 
in a cool or warm phase and is accreted 
before it has time to virialize.
Cool and warm gas would of course elude an X-ray 
investigation. Furthermore, {\it Chandra} observations 
do not have enough spatial resolution 
to investigate sources of gas located inside 
the sphere of influence of the SMBH (typically  
$\la 10$ pc). In this paper, we  
use optical brightness profiles 
to obtain a complementary estimate 
of the gas injection rate into the SMBH sphere 
of influence; from the stellar densities 
and population ages, we  
estimate the characteristic mass injection rates 
due to stellar outgassing.

\begin{deluxetable*}{lcccccccc}
\tabletypesize{\scriptsize}
\tablecaption{Targets of our study \label{tbl-1}}
\tablewidth{0pt}
\tablehead
{
\colhead{Galaxy} & \colhead{Type} & \colhead{$d$}
	& \colhead{$M_{\rm BH}$}
	& \colhead{$\log \dot{M}_{\rm B}$} 
	& \colhead{$\log \dot{M}_{\rm Edd}$}  
	& \colhead{$\log L_{\rm X}$} & \colhead{$\log L_{\rm Edd}$} 
	& \colhead{$(\dot{M}/\dot{M}_{\rm B})_{\rm A}$}\\[2pt]
   	& & \colhead{(Mpc)} & 
	\colhead{($10^8 M_{\odot}$)}& 
	\colhead{($M_{\odot}$ yr$^{-1}$)} & 
	\colhead{($M_{\odot}$ yr$^{-1}$)}&  \colhead{(erg s$^{-1}$)}
	& \colhead{(erg s$^{-1}$)} &  \\[2pt]
\colhead{(1)}  &  \colhead{(2)}  &  \colhead{(3)}  & \colhead{(4)}  &  
                  \colhead{(5)}  &  \colhead{(6)}  &  \colhead{(7)} 
		 &  \colhead{(8)} &  \colhead{(9)}}
\startdata
\, & \, & & & & & & & \\[6pt]
NGC\,821 
	& E6 & $24.1 \pm 2.0$ &  $0.85^{+0.35}_{-0.35}$ 
	& $-4.42$ & $0.29$ & $<38.7$ & $46.0$ & $<2$\\[4pt]
NGC\,3377 & E5-6& $11.2 \pm 0.5$ & $1.0^{+0.9}_{-0.1}$ 
	& $-4.56$ & $0.36$ & $38.5$ & $46.1$ & $1.3$\\[4pt]
NGC\,4486B & cE0 & $16.9 \pm 1.3$ & 
	$\left[6.0^{+3.0}_{-2.0}\right.$ 
	& $<-3$ & $1.14$ & $38.4$ & $46.9$& $\left.>0.1\right]$\\[4pt]
 	&&& $0.5^{+0.5}_{-0.2}$ & $<-5$& $0.06$& $38.4$ & 
	$45.8$ & $> 2$\\[4pt]
NGC\,4564 & E6/S0 & $15.0 \pm 1.2$  & $0.56^{+0.03}_{-0.08}$ 
	& $-5.27$ & $0.11$ & $38.9$&  $45.9$& $6$\\[4pt]
NGC\,4697 
	& E6 & $11.7 \pm 0.8$ &$1.7^{+0.2}_{-0.1}$ 
	& $-3.82$ & $0.59$ &$38.6$ &$46.3$ & $0.3$\\[4pt]
NGC\,5845 & E3 & $25.9 \pm 2.7$& $2.4^{+0.4}_{-1.4}$
	 & $-4.00$& $0.74$& $39.4$ & $46.4$ & $1.4$\\[4pt]
\enddata
\tablecomments{
Col.(1): galaxy ID.
Col.(2): from the NASA/IPAC Extragalactic Database (NED).
Col.(3): from Tonry et al.~(2001), except for NGC\,4486B,
	from Neilsen \& Tsvetanov (2000).
Col.(4): from Gebhardt et al.~(2003), except for NGC\,821 
        (Richstone et al.~2004), and for NGC\,4486B. 
	The latter has two alternative values. The first, 
	in square brackets,  
	is from Kormendy et al.~(1997), using a two-integral model.
	A similar result ($5.0^{+4.9}_{-4.8} M_{\odot}$) 
	was obtained with the same method by Kormendy \& Gebhardt
	(2001). 
	The alternative mass determination (a factor of $10$ lower) 
	was obtained in this paper (Section 2.2), from 
	the $M_{\rm BH}$--$n$ (S\'ersic index) relation 
	(Graham et al.\ 2003, 2005 in prep.).
Col.(5): Bondi inflow rate of the hot ISM into the SMBH 
	sphere of influence; see Paper I for details.
Col.(6): Eddington accretion rate, defined in Paper I.
Cols.(7),(8): unabsorbed X-ray luminosity of the nuclear source, 
	from {\it Chandra} observations (Paper I), and Eddington 
	luminosity.
Col.(9): accretion rates that would produce the observed 
	X-ray luminosities, at ADAF-like efficiencies, 
	normalised to the Bondi rate. The ADAF model 
        predicts $(\dot{M}/\dot{M}_{\rm B}) \sim 0.1$.
}
\end{deluxetable*}

We compare the observed X-ray luminosities of the nuclear 
sources with those expected from the total accretion 
rates: hence, we constrain the radiative 
efficiency and the fraction of gas that is 
really accreted, and discuss the possible fate 
of the gas component that does not 
reach the central BH. This allows us to 
discriminate between different accretion solutions 
which may be applicable to our systems.
Finally, we discuss the total accretion power 
budget and the condition for gas mass conservation 
inside the sphere of influence of the SMBH. 
We estimate the role of feedback-regulated outflows 
to achieve a mass equilibrium inside the sphere 
of influence, and we show that 
they are consistent with the 
energy budget of the system.


\section{Optical study}

\subsection{What we can learn from optical studies}

The main source of hot ISM in elliptical 
galaxies is thermalized gas 
lost by the stellar population, e.g., 
from red-giant winds (Ciotti et al.~1991).
Part of the gas that may be found in the nuclear region 
is produced locally; an additional contribution 
may come from a cooling flow, from larger radii.
Gas can also reach the inner galactic regions 
from the accretion of small satellites 
(e.g., Pipino et al.\ 2005); this is 
a fairly common process in ellipticals, 
and is thought to explain the origin 
of (some) nuclear disks.

Optical studies, with their better spatial resolution, 
can provide a better estimate of the stellar mass, 
and thus the stellar mass losses due to winds,
in the circumnuclear 
region, and therefore a better estimate of the 
gas available to fuel the SMBH. {\it HST} data 
were used for this purpose in Fabbiano et al.~(2004), 
to estimate the gas available in the nuclear region 
of NGC\,821. 
Inner brightness profiles also allow us 
to investigate the possibility
that some nuclei are X-ray faint because 
of a stellar deficit in the core, which 
may have resulted from the gravitational ejection
of stars during the coalesence of two or more SMBHs 
(Milosavljevic et al.\ 2002; Ravindranath, 
Ho \& Filippenko 2002; Graham 2004). 
In addition, using wider field-of-view 
ground-based images allows us to model 
the global light-profile and thereby acquire 
an alternative estimate of, and thus check on, 
the SMBH mass using the $M_{\rm BH}$--$n$ relation
(Graham et al.\ 2001, 2003; Novak, Faber \& Dekel 2005).

Another reason for studying the optical light-profiles 
is that the optical brightness can be used as a proxy 
for the stellar mass distribution, which 
in turn, for old populations, is proportional 
to the low-mass X-ray binary (LMXB) distribution 
(e.g., Kim \& Fabbiano 2003, 2004; 
Gilfanov 2004). Therefore, the global 
optical profiles offer a useful check 
on whether the unresolved X-ray 
emission is dominated by faint LMXBs or hot gas 
at large or small radii; and whether the X-ray emitting 
gas is more or less centrally concentrated than 
the X-ray binaries.

We have studied the optical light-profiles 
of the six galaxies in our sample.
Because of the large apparent (optical) extent of these galaxies on
the sky, we have taken the inner light-profiles from {\sl HST}
images and more extended profiles from ground-based images.  In the
case of NGC\,4564 we have been able to use the {\sl HST} light-profile
given in Trujillo et al.\ (2004) for both purposes.  Table~\ref{XXX}
provides the literature source for each light-profile.  In general,
major-axis $R$-band light-profiles have been used.  

\subsection{Global light-profiles and SMBH masses}

\begin{figure*} 
\includegraphics[angle=270,scale=0.7]{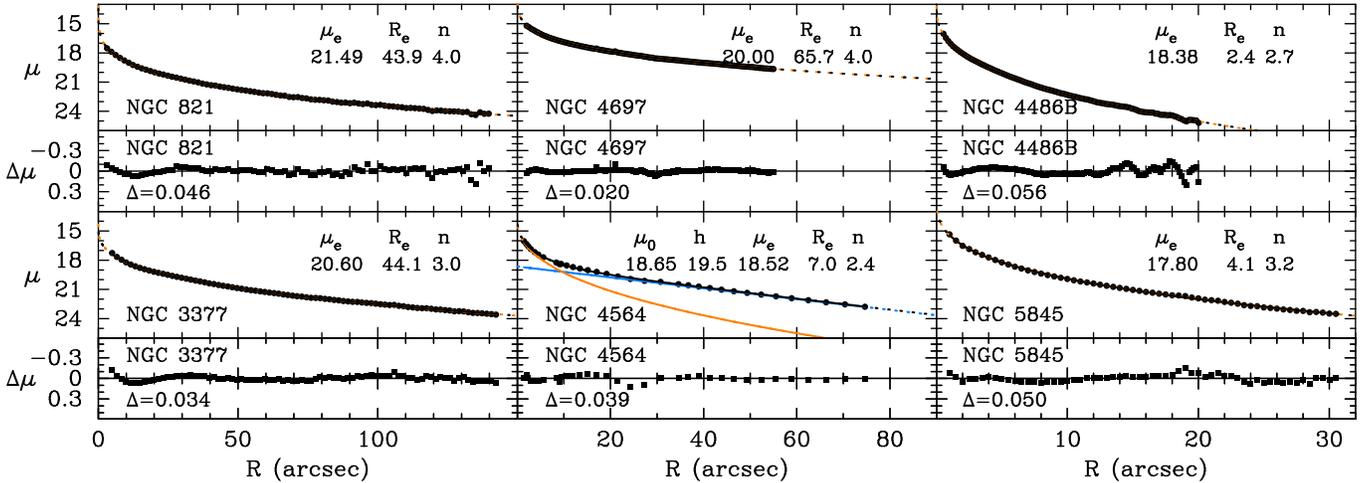}
\caption{
Major-axis light-profiles. The filters used are given in Table~\ref{XXX}.
The best-fitting S\'ersic models are shown, while a S\'ersic 
bulge $+$ exponential disk
decomposition is shown for NGC 4564. 
The residuals of the data about the model are shown beneath each galaxy. 
We used the S\'{e}rsic indices from these global fits 
as a check on the SMBH masses (Table 2), 
using the $M_{\rm BH}$--$n$ relation (Graham et al.\ 2001, 2003).
}
\label{YYY}
\end{figure*}

Figure~\ref{YYY} shows the best-fitting S\'ersic (1963, 1968) $R^{1/n}$
model parameters
for each galaxy's light-profile.
S\'ersic's $R^{1/n}$ model is often written as
\begin{equation}
I(R)=I_{\rm e} {\rm e}^{ -b_n\left[ (R/R_{\rm e}) ^{1/n} -1\right]},
\label{Sersic}
\end{equation}
where $I_{\rm e}$ is the (projected) intensity at the (projected)
effective radius $R_{\rm e}$.  The term $b_n$ is not a parameter but a
function of $n$ chosen in such a way that $R_{\rm e}$ encloses half of
the total light (Ciotti 1991; Caon et al.\ 1993; Graham \& Driver 2005).  
To minimize the influence of seeing, and the spurious 
effect on the global profiles that might have come from 
possible additional nuclear features or
depleted stellar cores, we have excluded 
the inner $\sim 1\arcsec$ of the profiles from the 
fitting process, which convolved the $R^{1/n}$ models with the
relevant point spread function before matching to the observed data.
Although generally classified 
as an elliptical galaxy, we concur with Trujillo et al.'s 
(2004, their Appendix C) re-designation
of NGC\,4564 as an S0 galaxy.  In addition to possessing a stellar ring
(Trujillo et al.\ 2004), its light-profile has the classic bulge plus
outer exponential disk structure (Figure 1).

We mentioned in Paper I that different kinematic SMBH mass 
determinations for NGC\,4486B were discrepant 
by more than one order of magnitude.
Moreover, Kormendy et al.\ (1997) report that the use of
anisotropic models reveals that NGC\,4486B 
may contain no SMBH at all.
To narrow down this uncertainty, we decided to use 
a secondary mass indicator for this galaxy. We estimated
its mass from the correlation between light-profile 
shape $n$ and nuclear BH mass $M_{\rm BH}$. 
Using $\log (M_{\rm BH}) = (3.02\pm 0.47)\log(n/3) 
+ (7.81\pm 0.07)$, with intrinsic variance 
$\epsilon = 0.27$ dex (Graham et al.\ 2005, in prep.),
we derive a mass of 
$0.5^{+0.5}_{-0.3} \times 10^8 M_{\odot}$, 
as given in Table~2 (Col.4). 
We compared this value with that expected 
from the velocity dispersion/SMBH mass correlation.
For a velocity dispersion
$\sigma = 170$ km s$^{-1}$ (from Hypercat), 
we expect $M_{\rm BH} \approx 0.6 \times 10^8 M_{\odot}$ 
if we use the correlation in Merritt \& Ferrarese (2001), 
or $M_{\rm BH} \approx 0.7 \times 10^8 M_{\odot}$ 
using the expression in Tremaine et al.~(2002). Similar values 
also follow from the bulge mass/BH mass correlation.
This agreement between all indirect indicators 
suggests that the kinematic mass measurement 
obtained from spherical, isotropic models 
(Kormendy et al.~1997)
may be an order of magnitude higher than the true value.
Throughout this paper, we tabulate  
both sets of values (kinematic measurement 
and $M_{\rm BH}$--$n$ value) for NGC\,4486B.


In order to check that the secondary mass indicators, 
in particular the $M_{\rm BH}$--$\sigma$
and $M_{\rm BH}$--$n$ relations, are reliable, 
we have tabulated (Table 2, Cols.4,6) 
the values of the SMBH masses estimated for the other five 
galaxies. 
In all cases, direct and indirect mass determinations 
agree within the errors, which strengthens our 
confidence in the indirect SMBH mass estimates 
for NGC,4486B, unless this is a really peculiar galaxy.
%


\subsection{Nuclear light-profiles}

We shall now study the stellar distribution 
in the nuclear region, which will provide 
an estimate of one source of gas available for accretion.
The cores of many luminous ($M_B \la -20.5$ mag) elliptical galaxies are
known to be partially depleted of stars (e.g., 
Quillen, Bower \& Stritzinger 2000; Rest et al.\ 2001;
Ravindranath et al.\ 2001; Trujillo et al.\ 2004, and references
therein).  In a hierarchical universe, giant elliptical 
galaxies are formed from the dry merger 
of smaller elliptical galaxies (Naab, Khochfar \& Burkert 
2005; Li, Mo \& van den Bosch 2005; De Lucia et al.\ 2005).  
The SMBHs from the
pre-merged galaxies sink to the center of the newly-wed galaxy where they
form a binary pair, the orbit of which decays by transferring orbital
angular momentum to nearby stars 
(Begelman, Blandford, \& Rees 1980).  Such a gravitational slingshot
evacuates the stars from the center of the new 
galaxy (Milosavljevic \& Merritt 2001).  A reduced
number of stars would imply a reduced local source of hot gas and may
explain the low X-ray luminosities.

\begin{figure*}
\includegraphics[angle=270,scale=0.7]{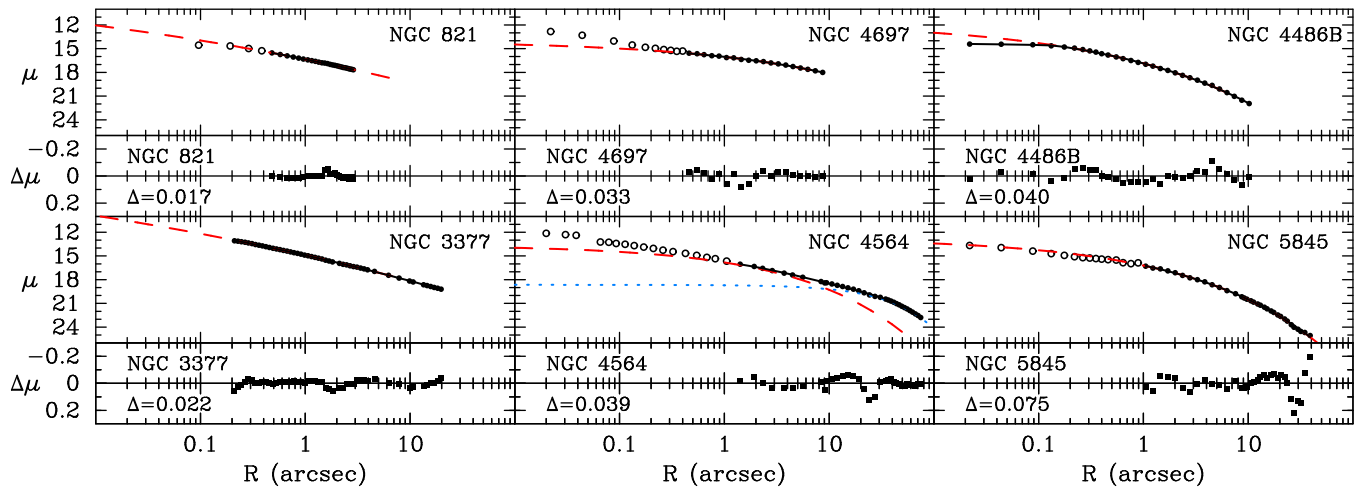}
\caption{
Major-axis, {\sl HST}-resolved, inner light-profiles.  The filters
used are given in Table~\ref{XXX}.  The long dashed curves show the
best-fitting (3-parameter) S\'ersic model.
For the lenticular galaxy NGC\,4564, an outer exponential disk
(short dashed line) has been included.  
For NGC\,4486B, the best-fitting core-S\'ersic model (Graham et al.\
2003) is also shown (solid curve).
Data points excluded from the fit are shown by the open circles (see
text for discussion).
}
\label{ZZZ}
\end{figure*}

The {\sl HST}-resolved, inner light-profiles are shown in
Figure~\ref{ZZZ}.  Again, the best-fitting S\'ersic model has been
shown, although this is only done here to help show deviations in the
data from the relatively smooth outer stellar distribution.  
%
NGC\,4486B is the only galaxy with an apparent core, rather than 
a ``power-law'' profile. We do note, however, that
even the short (300 s) {\sl HST} exposure of NGC\,3377 saturated at the
center.  A core may therefore exist inside of the innermost usable
data point at 0\farcs2.

The inner arcsec has been excluded from 
the S\'{e}rsic fit to several profiles.  NGC\,5845
has a prominent nuclear disk (Section 5.5; Quillen et al.\ 2000;
Trujillo et al.\ 2004) which has been excluded from the fit
shown in Figure~\ref{ZZZ}.  
NGC\,4697 also contains an additional nuclear feature,
probably a star cluster, that was excluded from the modeling of the
underlying host galaxy light, as done in Byun et al.~(1996). 
None of the inner light-profiles 
displays evidence of a central deficit of stars.
The apparent depletion in the core 
of NGC\,4486B is not due to a stellar deficit 
but to a double optical nucleus (Section 2.4).

Finally, we compared the optical light-profiles and diffuse X-ray 
surface brightness profiles (Figure 3) for the four 
galaxies in which we have sufficient counts 
in the X-ray emission. In two cases (NGC\,5845 and 
NGC\,4697) the optical profiles appear more centrally 
peaked. A detailed comparison and physical 
interpretation of the large-scale (a few kpc) 
optical and X-ray profiles is beyond the scope 
of this work. However, we have used the X-ray 
surface brightness profiles in the nuclear region 
to estimate the amount of hot gas available 
for accretion (Paper I); and we shall 
use the optical light-profiles to estimate the rate 
at which the SMBH can be fuelled by stellar winds 
(Section 2.5).
Taken together, these two components will give us 
a good estimate of the total rate at which gas 
reaches the sphere of influence of the SMBH, 
for the six galaxies in our {\it Chandra} sample.

\subsection{Peculiarities of NGC\,4486B}

Among the six galaxies in our sample, 
one stands out for its various peculiarities. 
NGC\,4486B is a rare compact elliptical (cE) galaxy.
The presence of a core in this galaxy (Figure 2)
contrasts with the cE galaxy M\,32, 
which has a steep inner power-law
that continues into the resolution limit of the data 
(Faber et al.~1997, their Fig.~2).
Although the inner light-profile for NGC\,4486B has come from an F555W
(roughly $V$-band) image, there is no sign of dust at its center.
The apparent stellar mass deficit arises from a lull in flux
between two off-centered optical nuclei, 
separated by $\approx 0\farcs15 \approx 12$ pc (Lauer et al.~1996; 
Kormendy et al.~1997). Thus, NGC\,4486B is the only other 
galaxy known so far with an optical double-nucleus 
around a single SMBH (the other galaxy is M\,31; 
Lauer et al.~1993).
The two optical nuclei have been interpreted 
as the pericenter and apocenter of an eccentric 
stellar disk around the SMBH (Tremaine 1995); 
such a disk might have resulted from the tidal disruption 
of a massive star cluster (Quillen \& Hubbard 2003).


The inner region of NGC\,4486B stands out in our sample 
also for its X-ray properties (Paper I). 
The nuclear X-ray source is point-like, 
with no detection of extended structures.
However, it is softer than typical AGN (Figure 3 and Table 2 
of Paper I), inconsistent with a $\Gamma \la 2.5$ power-law spectrum.
We do not have enough counts to determine 
whether its X-ray spectrum is a truly steep 
power-law, or a standard AGN spectrum with 
a soft excess (perhaps from dense, hot plasma).
As an aside, we note that a faint ($\sim 10^{38}$ erg s$^{-1}$), 
soft X-ray source was also detected at the optical 
position of the nucleus in the elliptical galaxies 
NGC\,4472 and NGC\,4649 (Soldatenkov, 
Vikhlinin, \& Pavlinsky 2003), although it is not clear 
what fraction of that emission comes from a central 
peak in the thermal plasma distribution and how much 
from the accreting SMBHs. And in M\,31, an (unrelated?) 
super-soft source is located at $\la 1\arcsec$ from 
the nucleus and might easily be confused for 
the accreting SMBH (Garcia et al.~2000) 
if that galaxy were at the distance of NGC\,4486B.
Given its physical peculiarities, 
and the large observational uncertainty 
in the mass of the SMBH, the gas density and the nature
of the nuclear X-ray source, we are aware 
that the accretion properties inferred for this 
galaxy may not be representative of typical 
quiescent elliptical galaxies.

\begin{deluxetable*}{lccccccccc}
\tabletypesize{\footnotesize}
\tablewidth{0pt}
\tablecaption{Optical surface brightness profiles, S\'{e}rsic 
indices and SMBH masses\label{XXX}}
\tablehead
{
\colhead{Galaxy} & \colhead{outer profile} & \colhead{$n$}  
	& \colhead{$M_{\rm BH,n}$}   
	& \colhead{$\sigma_{\rm e}$}
	& \colhead{$M_{\rm BH,\sigma}$}   
	& \colhead{$M_{\rm BH,kin}$}
        &           \colhead{inner profile} & \colhead{Type}  & \colhead{Notes}  \\[3pt]
\colhead{}       & \colhead{} & \colhead{}     
	& \colhead{($10^8 M_{\sun}$)} &
	\colhead{(km s$^{-1}$)} & \colhead{($10^8 M_{\sun}$)} &
         \colhead{($10^8 M_{\sun}$)} &          
	\colhead{} & \colhead{} &      \\[2pt]
\colhead{(1)}  &  \colhead{(2)}  &  \colhead{(3)}  & \colhead{(4)}  &  
                  \colhead{(5)}  &  \colhead{(6)} &  \colhead{(7)}
		&  \colhead{(8)}  &  \colhead{(9)} &  \colhead{(10)} 
}
\startdata
\, & \, & & & & & \\[6pt]
NGC 821   & R-band, 1 & $4.0 \pm 0.8$ 
	&  $1.5^{+2.2}_{-0.9}$ & $209$ & $1.6^{+0.4}_{-0.3}$  
       &  $0.85^{+0.35}_{-0.35}$
	 &  F160W, 5  & cusp & \\[4pt]
NGC 3377  & R-band, 1 & $3.0 \pm 0.6$ 
	&  $0.7^{+0.9}_{-0.4}$ & $145$ & $0.4^{+0.1}_{-0.1}$
	&  $1.0^{+0.9}_{-0.1}$ 
	&  F702W, 6  & cusp & 1 \\[4pt]
NGC 4486B & F555W, 2      & $2.7\pm0.5$
	 & $0.5^{+0.5}_{-0.2}$ & $170$ & $0.7^{+0.2}_{-0.2}$
	&  $\left[6.0^{+3.0}_{-2.0}\right]$ 
	&  F555W, 7  & ``dip'' & 2 \\[4pt]
NGC 4564  & F702W, 3  & $2.4^{+1.1}_{-0.4}$ 
	& $0.3^{+1.2}_{-0.2}$  & $162$ & $0.6^{+0.1}_{-0.1}$
	&  $0.56^{+0.03}_{-0.08}$
	&  F702W, 3  & cusp & 3 \\[4pt]
NGC 4697  & F475W, 4      & $4.0\pm0.8$ 
	&  $1.5^{+2.2}_{-0.9}$  & $177$ & $0.8^{+0.3}_{-0.2}$
	&  $1.7^{+0.2}_{-0.1}$
	&  F555W, 7  & cusp & 4,5 \\[4pt]
NGC 5845  & R-band, 1 & $3.2\pm0.6$ 
	& $0.8^{+1.1}_{-0.5}$ & $234$ & $2.5^{+0.7}_{-0.6}$
	&  $2.4^{+0.4}_{-1.4}$ 
	&  F702W, 3  & cusp & 1,5 \\[4pt]
\enddata
\tablecomments{
Col.(1): galaxy ID.
Col.(2): filter and source of the outer light-profile: 
         1 = Graham et al.\ (2001); 2 = {\it HST}/WFPC2 archive, 
		Proposal ID 6099, PI S.~Faber; 
         3 = Trujillo et al.\ (2004). 4 = {\it HST}/ACS archive, 
		Proposal ID 10003, PI C.~Sarazin.
Col.(3): best-fitting S\'ersic index $n$ for the outer brightness profile 
         (Figure~\ref{YYY}). 
Col.(4): predicted SMBH mass from the $M_{\rm BH}-n$ relation 
	(Graham et al.\ 2003; Graham et al.\ 2005, in prep.).
Col.(5): effective velocity dispersion, from Gebhardt et al.~(2003), 
	except for NGC\,4486B, for which we take the central velocity 
	dispersion from the Hypercat database. An alternative 
	measurement of the velocity dispersion in NGC\,4486B 
	is $\sigma \approx 185$ km s$^{-1}$ (Kormendy \& Gebhardt 2001) 
	which corresponds to a mass of 
	$1.0^{+0.3}_{-0.2} \times 10^8 M_{\odot}$. 
Col.(6): SMBH masses derived using the relation 
	in Tremaine et al.~(2002).
	Errors are derived assuming statistical plus 
	systematic uncertainty in the velocity 
	dispersion measurements $\approx 5\%$ (Gebhardt et al.~2003). 
Col.(7): SMBH masses derived from kinematic measurements, see Table 1 
	for references. 
Col.(8): filter and source of the inner brightness profile: 
         5 = Quillen et al.~(2000); 6 = Rest et al.\ (2001); 
         7 = Lauer et al.\ (1995); 3 = Trujillo et al.\ (2004). 
Col.(9): innermost (resolved) light-profile type. 
	The "dip" in the light-profile of NGC\,4486B is 
	created by its two optical 
	nuclei offset from the central position.
Col.(10): 1 = nuclear stellar disk; 2 = double optical nucleus, but a
	single central SMBH; 3 = galactic stellar disk; 
	4 = nuclear star cluster; 5 = dusty disk. 
}
\end{deluxetable*}


\begin{figure} 
\includegraphics[angle=270,scale=0.5]{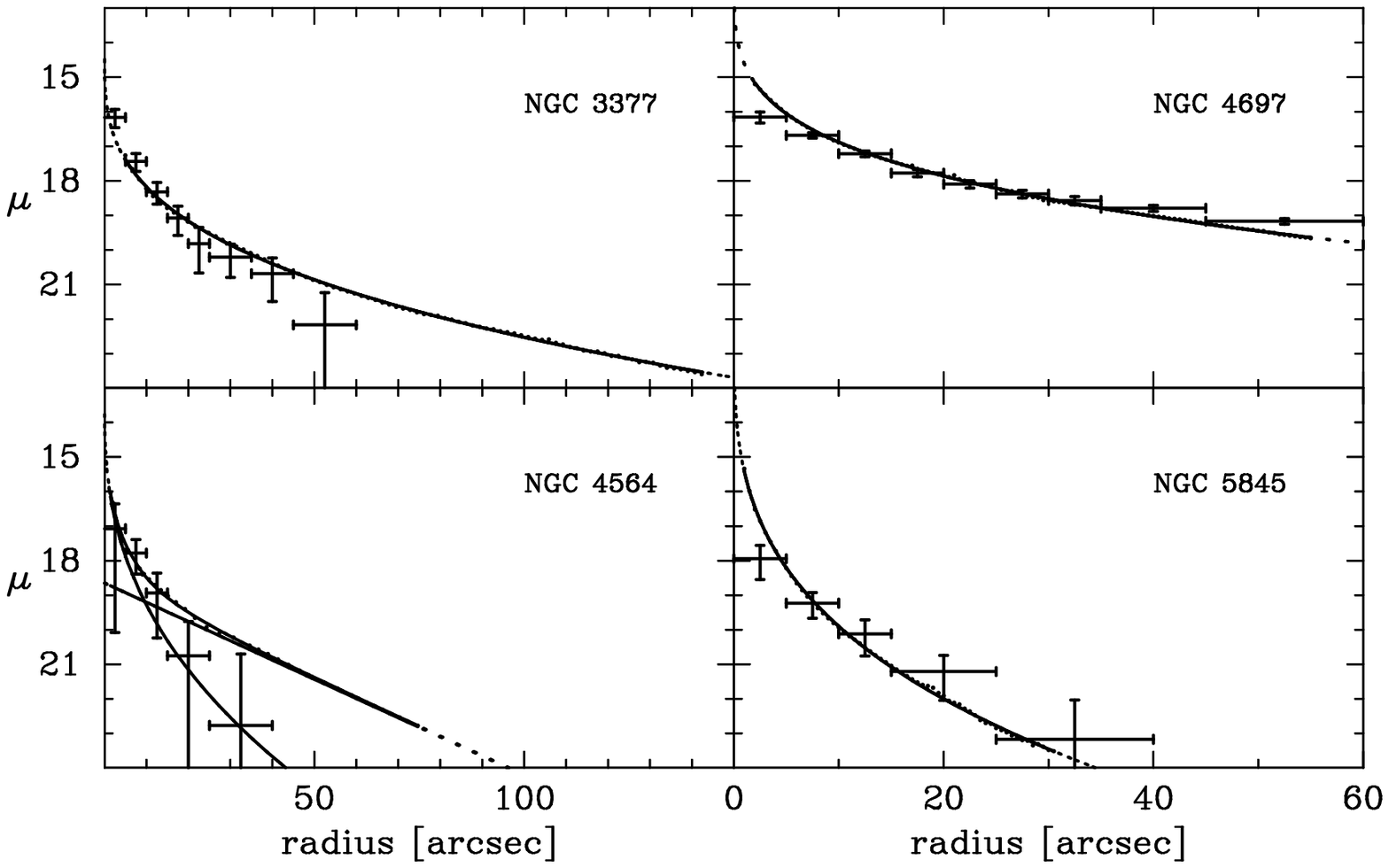}
\caption{
Comparison of the optical (curves) and soft X-ray 
(datapoints with error bars) brightness 
profiles. NGC\,4697 and NGC\,5845 have a relative central spike 
in the optical brightness profile (clearly visible 
also in Figures 5 and 6 of Paper I) which we cannot directly 
compare with the X-ray thermal-plasma emission, 
due to a lack of spatial resolution. This is one of the reasons 
why we need to take into account both the X-ray and 
the optical data to infer the gas injection rate.
In the S0 galaxy NGC\,4564, the radial profile of 
the diffuse X-ray emission seems to follow more 
closely the optical brightness profile of the virialized 
stellar component (bulge), 
rather than that of the exponential disk.
}
\label{fig-17}
\end{figure}



\subsection{Stellar mass-loss rates} 

An estimate for the stellar mass-loss 
contribution to the ISM comes from Ciotti et al.~(1991) 
and is given by:
\begin{equation}
\dot{M}_{\ast}(t) \approx 1.5 \times 10^{-11} 
	\left(\frac{L_{B}}{L_{\odot, B}}\right) 
	\left(\frac{t}{15 {\rm ~Gyr}}\right)^{-1.3} 
	\ M_{\odot} {\rm~yr}^{-1},
\end{equation}
where $\dot{M}_{\ast}(t)$ is the rate at which 
a stellar population of age $t$ and $B$-band luminosity 
$L_B$ loses gas through stellar winds.
Mass is also injected into the ISM 
by Type-Ia SNe, at a rate which depends 
on specific SN models, and is also 
proportional to $L_B$, but is only 
a few percent of $\dot{M}_{\ast}$ (Ciotti et al.\ 1991, 
their Eq.~7). We shall henceforth consider this 
component to be included in $\dot{M}_{\ast}$.
In quasi-stationary conditions, we expect 
this gas to flow towards the galactic center 
and to be shock-heated or virialized to temperatures 
$kT \sim 0.5$ keV, giving rise to the hot ISM 
seen, for example, in five of the six galaxies 
of our sample. For a detailed study 
of the physical processes and timescales 
involved in the conversion of gas from the warm 
to the hot phase, see Mathews (1990). 

As a first approximation, 
we shall take the Bondi rate $\dot{M}_{\rm B}$ 
(Table 1; definition and discussion in Paper I, Section 4.3)
to represent the rate of inflow of diffuse hot gas into 
the sphere of influence of the SMBH 
(i.e., across the surface of radius $r_{\rm acc}$); 
we shall assume 
that this term takes into account the whole stellar 
contribution from outside $r_{\rm acc}$.
A fraction $\la 1$ of this inflowing gas will 
then sink into the SMBH; the rest may build up 
inside the nuclear region, or be ejected in a wind.
In addition, we need to estimate how much 
gas is produced from stellar winds by 
the stellar population 
directly inside $r_{\rm acc}$. 

To obtain an estimate of $L_B$ inside the SMBH sphere of influence 
we need to integrate the optical-luminosity density profiles 
over the sphere of radius $r_{\rm acc}$. 
Firstly, we used the luminosity density profiles
plotted by Gebhardt et al.~(2003) 
(taking into account the $B-V$ color index 
for each galaxy, from Faber et al.\ 1997 and 
the Third Reference Catalogue of Bright Galaxies, 
de Vaucouleurs et al.~1991).
The only galaxy in our sample for which 
the light-profile is not available 
in Gebhardt et al.~(2003) is NGC\,4486B.
We have also performed our own
deprojection of the light-profiles 
for all six of our target galaxies, using 
the fitted S\'ersic models\footnote{
In three of the six galaxies, 
NGC\,4486B, NGC\,4564, and NGC\,4697, 
the S\'{e}rsic models shown in Figures 1 and 2 
provide a good fit to the underlying galaxy light,  
provided that the additional nuclear component 
in the innermost $\approx 1\arcsec$ (disk 
or star cluster) is excluded from the fit. 
While this is the standard procedure for the 
correct determination of $n$ (Table 2), 
it would lead to an under-estimation 
of the total flux from inside the sphere of influence.
So, for the purpose of determining $L_B$, 
we refitted new S\'{e}rsic models to the very inner 
light-profiles of those three galaxies, 
including also the additional nuclear components.}
(see Graham \& Colless 1997, their Appendix B).
We note that a potential complication
may arise from our deprojection procedure 
which assumes a spherical stellar distribution.  
The presence of flat nuclear disks clearly
depart from this assumption of sphericity.  
Nonetheless, our estimated volumetric luminosity 
$L_B$ inside of $r_{\rm acc}$ agrees with that
from Gebhardt et al. (2003) to within 
a factor of two for every galaxy, after correcting 
for the different color bands. 
Finally, for each galaxy, we have taken 
the average between the two values of the integrated $L_B$, 
obtained from Gebhardt's (2003) profiles and our own 
analysis (except for NGC\,4486B, 
where only our value is available). These values were used 
in equation (2) to determine the mass loss rates.

Assuming an old population (12 Gyr) for all galaxies, 
we can estimate the mass loss rates 
within the accretion radius of each galaxy 
($\sim 10$ pc, Table 7 in Paper I); typical values are 
$\dot{M}_{\ast} \sim 10^{-4} M_{\odot}$ yr$^{-1}$ (Table 3, Col.3)
(see also Fabbiano et al.~2004 for NGC\,821).
In fact, these are lower limits: there is evidence 
that the stellar population in the nuclear region 
is younger for all six galaxies, ranging 
from $\approx 2.5$ to $\approx 10$ Gyr (Table 3, Col.2); 
taking the younger age into account yields 
higher mass loss rates (Table 3, Col.4).
As we shall discuss in Section 4.1, a fraction $\la 1$ 
of this gas sinks into the BH, the rest may be ejected 
in a galactic wind, or settle into a nuclear disk, or be 
recycled into other stars. 

In summary, so far we have checked the masses 
of the SMBHs and estimated the rates 
at which warm and hot (Paper I) gas 
becomes potentially available for accretion, 
inside their spheres of influence.
In the next section, we shall compare these rates 
with the rates at which gas actually accretes, 
indirectly estimated 
from the observed X-ray luminosities.


\section{Where does the accretion power go?}

\subsection{Accretion rates and efficiencies}

We define a total accretion power
$P_{\rm acc} \equiv \eta' \dot{M} c^2$, 
where $\eta'$ is the total accretion 
efficiency (radiative plus mechanical), and 
$\dot{M}$ the rate at which matter actually 
accretes onto the SMBH. $\dot{M}$ receives contributions 
from both the hot gas captured at the Bondi radius 
(Table 1 and Paper I), and the additional 
gas injection from stellar-mass losses (Eq.[2] and Table 3).
In general, $\dot{M} \le (\dot{M}_{\rm B} + \dot{M}_{\ast})$
because only a fraction of the available gas 
may reach the BH; different accretion models 
predict different values for this fraction.

Similarly, we define a bolometric luminosity  
$L_{\rm bol} = \eta \dot{M} c^2$, where 
the radiative efficiency $\eta \equiv f_{\rm r}\eta' \la \eta'$ 
(Paper I, Section 5.1).
For a standard disk, $\eta' \sim \eta \sim 0.1$; 
for purely advective accretion, $\eta \la \eta' \ll 0.1$; 
in a radiatively-inefficient system dominated 
by a fast jet or a massive outflow, $\eta' \sim 0.1$ and 
$f_{\rm r} \ll 1$.
The X-ray luminosity in the $0.3$--$10$ keV band 
is $L_{\rm X} \equiv f_{\rm X} L_{\rm bol} \sim 0.1 L_{\rm bol}$
(Elvis et al.~1994; Ho 1999).
Finally, we shall use dimensionless accretion rates 
$\dot{m}\equiv 0.1 \dot{M} c^2/L_{\rm{Edd}}$,  
$\dot{m}_{\rm B} \equiv 0.1 \dot{M}_{\rm B} c^2/L_{\rm{Edd}}$ 
and $\dot{m}_{\rm t} \equiv 
0.1 (\dot{M}_{\rm B}+ \dot{M}_{\ast})c^2/L_{\rm{Edd}}$.

We found in Paper I that, for the SMBHs in 
our six target galaxies,
$L_{\rm X}/L_{\rm{Edd}}  \sim 10^{-8}$--$10^{-7}$, 
and that the dimensionless Bondi inflow rates of the hot ISM 
$\dot{m}_{\rm B}  \sim 10^{-5}$ 
(Table 1, and Figure 14 in Paper I). 
But the hot ISM inflow is only a lower 
limit to the total gas injection rate.
From our estimate of the stellar mass losses 
inside $r_{\rm acc}$ (Section 2.5, and Table 3, Col.9), 
we find that the stellar-wind contribution 
is higher than the Bondi inflow rate of hot gas  
for our galaxy sample 
(all of which have very low Bondi rates); 
this was already pointed out for NGC\,821 
in Fabbiano et al.~2004. 
Hence, we estimate that  
the dimensionless, total gas injection rate 
is $\dot{m}_{\rm t}  \sim 10^{-4}$--$10^{-3}$ 
(Table 3). 

In Paper I, we compared the observed X-ray luminosities 
of the SMBHs with the inferred Bondi inflow rates, 
and found no clear correlations, as previously noted 
by Pellegrini (2005).
In view of the results of our optical study, 
we now suggest that the physical interpretation 
of the faint SMBH emission becomes clearer 
if we compare the X-ray luminosity 
with the total gas injection instead, wherever possible:
\begin{equation}
\frac{L_{\rm{X}}}{0.1\left(\dot{M}_{\rm{B}}+\dot{M}_{\ast}\right)c^2} 
	= f_{\rm X} f_{\rm r} \left(\frac{\eta'}{0.1}\right) 
	\left(\frac{\dot{m}}{\dot{m}_{\rm t}}\right) 
	\sim 10^{-5}{\rm \,-\,}10^{-4}.
\end{equation}
So, the quiescent 
SMBH luminosity can be explained by a combination
of low efficiency and low accretion fraction, such that 
$\eta (\dot{m}/\dot{m}_{\rm t}) \sim 10^{-5}$--$10^{-4}$.
We discuss now which accretion solutions may  
satisfy this constraint.

\subsection{Advective accretion (ADAF models)}

We showed in Paper I that Advection-Dominated 
Accretion Flows (ADAF, 
Narayan \& Yi 1995; Narayan 2002) under-predict the X-ray 
luminosities of most SMBHs with Bondi inflow 
rates $\dot{m}_{\rm B} \la 10^{-4}$, but over-predict 
the X-ray luminosities of most SMBHs with 
$\dot{m}_{\rm B} \ga 10^{-4}$.
We shall now compare the observationally-determined 
(unabsorbed) X-ray luminosities with the total gas injection 
rates $\dot{m}_{\rm t}$.
Firstly, we recall that in the self-similar ADAF model, 
the luminosity  
$L_{\rm ADAF} \approx 0.1\dot{M}c^2(\dot{m}/\alpha^2) 
\approx (\dot{m}/\alpha)^2 L_{\rm Edd}$ 
(Narayan \& Yi 1994), 
where the viscosity parameter $\alpha \sim 0.1$;  
the radiative efficiency 
$\eta \sim 0.1 \dot{m}/\alpha^2 \sim 10 \dot{m}$. 
From a more rigorous calculation 
(Merloni, Heinz \& Di Matteo 2003), 
the X-ray luminosity $L_{\rm X,\,ADAF} 
\propto \dot{m}^{2.3} L_{\rm Edd}^{0.97}$. 
If all the gas is accreted 
onto the SMBH, {\em even at ADAF-like efficiencies}, 
it would produce X-ray luminosities much higher than 
observed (Table 3, Cols.6 and 7).
Conversely, we have estimated the fraction of gas 
available that has to be accreted 
at ADAF-like efficiencies, to produce the observed 
luminosities: we obtain $\dot{m}/\dot{m}_{\rm t} \sim 0.01$--$0.1$ 
for the galaxies in our sample (Table 3, Col.10).

We have already noted (Paper I) that most SMBHs with Bondi rates 
$\dot{m}_{\rm B} \ga$ a few $10^{-4}$ are consistent  
with radiatively-inefficient accretion 
at a rate $\dot{m}/\dot{m}_{\rm B} \la 0.1$; 
this obviously implies also $\dot{m}/\dot{m}_{\rm t} \la 0.1$.
We speculate that accretion onto 
this latter group of SMBHs is dominated 
by the hot-ISM inflow, i.e., $\dot{m}_{\rm t} \sim 
\dot{m}_{\rm B} \ga \dot{m}_{\ast}$ when $\dot{m}_{\rm B} 
\ga$ a few $10^{-4}$. In the sample of early-type 
galaxies for which luminosity density profiles 
are available (Gebhardt et al.~2003, and our own analysis 
in Section 2.5), 
the integrated $B$-band luminosities inside 
the accretion radii correspond to mass loss 
rates $\dot{m}_{\ast} \la$ a few $10^{-4}$, 
for an old or intermediate-age population.
Estimating the total accretion rate for all the sources 
in Pellegrini's (2005) sample is left 
to further work.

We conclude that radiatively-inefficient models 
can explain the observed luminosity 
of our target SMBHs, with accretion fraction 
$\la 10\%$ of the total gas injected 
into the sphere of influence of the SMBH.
Hence, the puzzling discrepancy between two 
groups of ``underluminous'' and ``overluminous'' SMBHs 
(with respect to the ADAF predictions) 
is naturally resolved, when we take into 
account all the sources of fuel.
We shall discuss the relation between 
SMBH luminosity and accretion rate 
in a more quantitative way in Section 4, 
in the framework of a simple phenomenological model 
which will describe the accretion power and mass budget. 

In fact, in a few cases the accretion fraction 
$\dot{m}/\dot{m}_{\rm t} \ll 10\%$.
For example, NGC\,4649 has a dense hot-ISM 
in the central region: $n_e \approx 0.5$ 
cm$^{-3}$, corresponding to a Bondi inflow 
rate $\dot{m}_{\rm B} \approx 10^{-2.5}$
(Pellegrini 2005; Soldatenkov et al.~2003).
The stellar mass loss rate within the sphere of influence 
is much smaller, $\dot{m}_{\ast} \approx 10^{-8}$, 
due to the low surface brightness and stellar density 
in the inner region (Gebhardt et al.~2003).
The observed X-ray luminosity of the nuclear 
source is $\approx 10^{38}$ erg s$^{-1}$ 
(Soldatenkov et al.~2003); at an ADAF-like efficiency, 
this implies an accretion fraction 
$\dot{m}/\dot{m}_{\rm t} \sim 10^{-3}$.
For this object, radiatively-inefficient solutions 
based on convection or outflows (e.g., ADIOS and CDAF models) 
are strongly favoured over the basic ADAF scenario.

\subsection{Advection, convection, outflows or jets?}

Our combined X-ray and optical analysis 
of quiescent SMBHs has suggested 
that only a fraction $\la 10\%$ of the total gas available 
from stellar winds and hot-ISM Bondi inflow 
is accreted onto the compact object, {\it and} this process 
is radiatively-inefficient, that is $\eta < 0.1$.
Hence, we must now debate what happens to the rest 
of the energy potentially available from accretion, 
and what happens to the gas that 
does not get accreted.

In the original ADAF scenario, all the energy 
that is not radiated simply disappears into the SMBH. 
In other varieties of advective flows, for example  
Convection Dominated Accretion Flows (CDAFs: 
Narayan, Igumenshchev, \& Abramowicz~2000), 
or Advection Dominated Inflow-Outflow Solutions 
(ADIOS: Blandford \& Begelman 1999), 
part of the accretion power is used to 
sustain convection or slow, massive outflows. This 
scenario has the advantage of providing 
at the same time a physical mechanism 
for reducing the accretion rate, preventing 
most of the gas from reaching the SMBH.
We shall discuss the role of outflows in Section 5, 
when we outline the conditions for mass equilibrium 
inside the sphere of influence.

Another possibility is that most 
of the non-radiative accretion power 
is carried outwards as mechanical 
luminosity in a fast jet.
Models based on a combination of 
radiatively-inefficient accretion flows plus powerful 
jets ($P_{\rm J} \sim 0.1 \dot{M} c^2$) 
can explain the total energy balance
in the case of other low-luminosity ellipticals such as M\,87
(Di Matteo et al.~2003), IC\,4246 (Pellegrini et al.~2003) 
and IC\,1459 (Fabbiano et al.~2003).

It was recently suggested that the dichotomy between 
high-luminosity and low-luminosity AGN corresponds 
to two different accretion modes (Jester 2005, 
and references therein): thermally dominated 
(high power from a radiatively efficient disk) 
and non-thermally dominated (low power from a radiatively 
inefficient flow), respectively. In the latter class 
of sources, the non-thermal emission may come entirely 
from a steady, compact jet (e.g., Falcke, 
K\"{o}rding \& Markoff~2004 
and references therein); this state corresponds 
to the ``low/hard'' state of stellar-mass X-ray binaries. 
The transition between the two 
modes is expected to take place at $\approx$ a few percent 
of the Eddington luminosity (Fender, Belloni \& Gallo~2004), 
a threshold above which a standard thin disk 
is formed and the jet is quenched.
Alternatively, this transition could be due to the change in
the accretion rate caused by an ionization instability 
(Janiuk, Siemiginowska, \& Szczerba 2004; Siemiginowska et al.~1996).

Significant support to this scenario has come 
from the ``fundamental plane'' correlation 
found by Merloni et al.~(2003) 
between X-ray and radio luminosity and BH mass 
in a sample of $\sim 100$ SMBHs (mostly low-luminosity AGN) 
and a few stellar-mass X-ray binaries.
None of the six galaxies in our sample have 
significant radio detections; however, five have
at least reliable upper limits to their radio-core emission, 
from VLA observations (Section 3.4 in Paper I).
Combining the ``core'' X-ray luminosities with 
the radio flux limits, we note that four of the SMBHs are 
radio faint, falling below the best-fit empirical correlation
of Merloni et al.~(2003) (Figure 4). However, the relation has a large 
scatter ($\sigma_{\log R} = 0.88$ dex), 
and our target sample was selected to be radio faint. 
Deeper radio observations are needed to determine 
how discrepant these four SMBHs are from those used to derive 
the fundamental plane correlation, and 
whether or not they belong to a different, radio-quiet 
population without a steady jet, 
perhaps dominated by advection or by slow, massive 
outflows instead.

We have also determined the radio-loudness parameter 
$R_{\rm X} \equiv \nu L_{\nu}/L_{\rm X}$ 
(Terashima \& Wilson 2003), which is a ratio 
of radio core luminosity at 5 GHz over nuclear 
X-ray luminosity 
in the $2$--$10$ keV band, useful for the classification 
of AGN and quasars. We obtain $R<-3.1$ (NGC\,3377), 
$R<-2.6$ (NGC\,4564), $R<-3.2$ (NGC\,4697), 
and $R<-3.3$ (NGC\,5845). These upper limits
are consistent with typical ratios expected from 
low-luminosity AGN (Fig.~4 in Terashima \& Wilson 2003), 
extrapolating to luminosities in the 
$\sim 10^{38}$--$10^{39}$ erg s$^{-1}$ range. 
The conventional boundary between 
radio-loud and radio-quiet nuclei 
is at $R = -4.5$, with all low-luminosities 
AGN being radio-loud. Again, radio observations 
of our targets about one of order of magnitude deeper 
will determine whether they belong to a new kind 
of X-ray faint, radio quiet SMBH population---or, at least, 
less radio loud than the low-luminosity AGN 
in Terashima \& Wilson's (2003) sample.

The morphologies of the X-ray emission 
in the nuclear regions suggest possible 
jet-like features in two of our target galaxies, 
NGC\,821 and NGC\,3377. The former was discussed 
in Fabbiano et al.~(2004), who suggest 
that the X-ray flux is consistent with synchrotron
emission in a jet, or with hot thermal plasma 
shocked by intermittent nuclear outbursts.
For NGC\,3377, see Paper I, Figure 7.


\begin{figure}
\includegraphics[angle=270,scale=0.36]{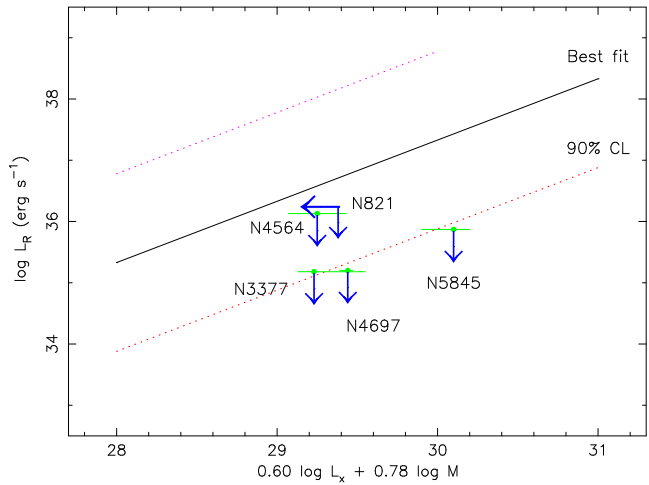}
\caption{Location of five of our target galaxies 
(those for which an upper limit to the radio flux 
$L_{\rm R}$ is available) in the fundamental plane of BH activity  
The best-fit line and the $90\%$ confidence limit 
are taken from Merloni et al.~(2003). The radio flux 
limits are listed in Paper I, Section 3.4. SMBH masses 
and observed (de-absorbed) X-ray luminosities 
are listed in Table 1.
\label{fig11}}
\end{figure}


\section{Balancing the mass budget}

\subsection{Disposing of the excess gas}

Since the inflow of gas within the accretion 
radius seems to be much higher 
than what sinks into the SMBH, 
we want to determine whether the gas  
builds up until it reaches an equilibrium 
between injection and depletion (accretion 
through the event horizon, plus outflows 
and any other sink terms), and what the equilibrium 
level is; or whether the gas may be depleted 
in intermittent outbursts without 
reaching a steady state.

Mass outflows are the most natural mechanism  
for disposing of the excess gas and achieving a mass 
equilibrium (e.g., Di Matteo et al.~2000). 
We shall determine how much 
mass has to be carried outwards in an outflow 
to ensure a mass balance, how much 
power is required, 
and what source of energy may power the outflow.
Although we shall focus on outflows 
for this work, we note that scenarios 
based on star formation in a nuclear disk 
or nuclear star cluster 
could also in principle provide an outlet 
for the excess gas, without the need for 
outflows. NGC\,5845 could be one such 
example, with its dusty stellar disk.
However, the evolutionary timescale 
and mass deposition rate in nuclear disks 
are at present unknown (Kormendy et al.~2005).

In the nuclear star-formation scenario, 
sequestration of cold gas into stars 
could be responsible for the low accretion 
rate onto the SMBH. An observational prediction of this model 
is the presence of a thin disk of dust and gas (including 
cold molecular gas) in the nuclear region, 
with ongoing star-formation; this would be detectable, 
although not necessarily resolved, 
from the usual indicators, such as H$\alpha$ 
and far-IR emission, and an X-ray emission component 
from SN remnants and high-mass X-ray binaries 
(in addition to the nuclear X-ray source). 
The characteristic size of such disks should be 
comparable with the Bondi accretion radius 
(Paper I, Section 4.3)
that is $\la$ a few $10$s pc; their radiative 
spectrum would peak in the mid- or 
far-IR wavelength range.
This scenario was found to be consistent 
with the spectral energy distribution of M\,87 
(Tan \& Blackman 2004), and could also be at work 
in some X-ray faint ellipticals.
The far-IR, rather than the X-ray, is the most 
suitable energy band for such studies; we have 
obtained time on {\it Spitzer} to carry out 
such an investigation.


\begin{deluxetable*}{lccccccccc}
\tabletypesize{\footnotesize}
\tablewidth{0pt}
\tablecaption{Stellar mass-loss rates, total accretion rates and 
	X-ray luminosities (expected and observed)\label{XXY}}
\tablehead
{\colhead{\,}&\colhead{\ }&\colhead{}&\colhead{}
	&\colhead{\ }&\colhead{}&\colhead{} &&&\\[8pt]
\colhead{Galaxy} & \colhead{Age} & \colhead{$\dot{M}_{\ast,1}$}  
& \colhead{$\dot{M}_{\ast,2}$} 
& \colhead{$\dot{M}_{\ast,2}/\dot{M}_{\rm Edd}$}
& \colhead{$\log (0.1\dot{M}_{\ast,2}c^2)$} 
& \colhead{$\log L_{\rm{X, \, ADAF}}$}
& \colhead{$\log L_{\rm{X, \, obs}}$}
& \colhead{$\dot{M}_{\ast,2}/\dot{M}_{\rm B}$}
	&\colhead{$\dot{M}/\dot{M}_{\rm t}$}\\[2pt]
\colhead{}       & \colhead{(Gyr)} & \colhead{($M_{\odot}$ yr$^{-1}$)} &
                   \colhead{($M_{\odot}$ yr$^{-1}$)} 
	& &  (erg s$^{-1}$)
	& (erg s$^{-1}$)  & (erg s$^{-1}$)
	&   & \\[2pt]
\colhead{(1)}  &  \colhead{(2)}  &  \colhead{(3)}  & \colhead{(4)}  &  
                  \colhead{(5)}  &  \colhead{(6)}  &  \colhead{(7)} 
		 &  \colhead{(8)} &  \colhead{(9)} &  \colhead{(10)}}
\startdata
\, & \, & & & & & & & & \\[6pt]
NGC\,821   & $4.0^{+1.7}_{-3.5}$ & $1.7\times 10^{-4}$
	& $7.1\times 10^{-4}$ & $3.6\times 10^{-4}$ & 
	$42.6$ & $41.3$ 
	& $<38.7$ & $18.7$ &$<7\%$\\[4pt]
NGC\,3377  & $3.7^{+0.4}_{-0.5}$ & $5.5\times 10^{-4}$
	& $2.6\times 10^{-3}$ & $1.1\times 10^{-3}$ & 
	$43.2$  & $42.5$
	& $38.5$ & $92.9$ & $2\%$\\[4pt]
NGC\,4486B & $9.5^{+1.5}_{-1.5}$ & $\left[1.3\times 10^{-4}\right.$ 
	&  $1.8\times 10^{-4}$   & $1.2 \times 10^{-5}$ & $42.0$   
	&  $39.8$  &   $38.4$& (NA) & $5$--$\left.25\%\right]$\\[4pt]
	& & $1.4\times 10^{-5}$ 
	&  $1.9\times10^{-5}$   & $1.6 \times 10^{-5}$ & $41.0$   
	&   $39.0$  &   $38.4$& (NA) & $45$--$55\%$\\[4pt] 
NGC\,4564  &  $\approx 8$ & $1.7\times 10^{-4}$
	& $4.0\times 10^{-4}$ & $3.1\times 10^{-4}$ & 
	$42.4$  & $41.0$
	& $38.9$ & $74.1$ & $12\%$\\[4pt]
NGC\,4697  & $8.9^{+1.9}_{-1.9}$ &  $5.7\times 10^{-4}$
	& $8.5\times 10^{-4}$ & $2.2\times 10^{-4}$ & 
	$42.7$ & $41.2$
	& $38.6$ & $5.7$ & $7\%$\\[4pt]
NGC\,5845  & $2.5^{+0.4}_{-0.5}$ & $4.0\times 10^{-4}$
	& $3.1\times 10^{-3}$ & $5.7\times 10^{-4}$ & 
	$43.2$ & $42.1$
	& $39.4$ & $31.3$ & $7\%$\\[4pt]
\enddata
\tablecomments{
Col.(1): galaxy ID. For NGC\,4486B, the bracketed line 
	refers to the higher mass estimate (see Table 1).
Col.(2): age of the stellar population in NGC\,821, NGC\,3377 and 
	NGC\,5845 from Denicol\'{o} et al.~(2005); age in NGC\,4697 
	from Trager et al.~(2000); age in NGC\,4564 from 
	Sil'chenko (1997); age in 
	NGC\,4486B from Sanchez-Blazquez (2004). 
Col.(3): mass loss rates from Ciotti et al.~(1991), assuming 
	an age of 12 Gyr.
Col.(4): mass loss rates from Ciotti et al.~(1991), assuming 
	the ages listed in Col.(2).
Col.(5): normalized stellar mass loss rate.
Col.(6): expected SMBH luminosity if all the gas injected by the stellar 
	population inside the sphere of influence is accreted 
	with standard radiative efficiency ($\eta = 0.1$). 
Col.(7): X-ray luminosity predicted by the ADAF solution, 
	from the model fits of Merloni et al.~(2003), 
	assuming $\dot{M} = \dot{M}_{\ast,2}$, and viscosity 
	parameter $\alpha = 0.1$.
Col.(8): observed X-ray luminosity from the nuclear sources, from 
	Paper I.
Col.(9): comparison between the  
	gas injection rate into the sphere of influence  
	from stellar mass losses and from Bondi inflow of the hot ISM.
	(No diffuse hot gas is detected in NGC\,4486B, see Paper I.)
Col.(10): fraction of the (total) injected gas that has to be accreted 
	onto the SMBH, to reproduce the observed 
	luminosities at an ADAF-like radiative efficiency 
	(i.e., $\eta \sim 10\dot{m}$; more precise values obtained 
	from the model fit of Merloni et al.~2003).
}
\end{deluxetable*}


\subsection{Outflows and mass equilibrium: a phenomenological model}

We shall indicate with $M_{\rm a}(t)$ 
the total gas mass inside the SMBH sphere of influence 
(radius $r_{\rm acc}$) at a time $t$;
with $\dot{M}_{\rm w}(t)$, the net outflow rate 
from the same region; with $\dot{M}_{\rm s}$, 
the star plus dust formation rate 
inside that region;
with $\dot{M}_{\ast}$ and 
$\dot{M}_{\rm B}$, the gas injection rate 
from the stellar population 
and the hot ISM respectively.
$\dot{M}_{\ast}$ and $\dot{M}_{\rm B}$ depend 
on the stellar and gas content of the galaxy: 
they vary over the Hubble timescale 
but can be taken as approximately constant  
over timescales $\la 10^8$ yr, 
i.e., shorter that the evolutionary timescale 
of the stellar population and the cooling time 
of the X-ray emitting gas. For quasi-spherical inflows, 
the accretion timescale $\la 10^6$ yr, which is also 
much shorter than the timescale considered here.
The total mass $M_{\rm a}$ 
is obtained by solving the equation
\begin{equation}
\dot{M}_{\rm a}(t) =  \dot{M}_{\rm B} +  \dot{M}_{\ast} 
         - \dot{M}(t) - \dot{M}_{\rm w}(t) -\dot{M}_{\rm s}(t), 
\end{equation}
where the accretion rate onto the SMBH, $\dot{M}$, 
contains contributions from both the inflowing hot ISM 
and the stellar mass losses inside $r_{\rm acc}$:
\begin{equation}
\dot{M} = a \dot{M}_{\rm B} +  b \dot{M}_{\ast},
\end{equation}
with $a \le 1$ and $b \le 1$. 
In principle, $a$ and $b$ may be a 
function of time and other parameters of the system, 
such as the gas density and temperature.
Both from a comparison with the observed 
X-ray luminosities (Table 3) and from theoretical 
arguments (briefly recalled in Paper I), 
we expect that $a \sim b \la 0.1$ for a radiatively-inefficient 
inflow (ADAF, CDAF or ADIOS).
Either $\dot{M}_{\rm B}$ or $\dot{M}_{\ast}$ 
may be the dominant term in the various galaxies, 
depending on the stellar density and age, and on the 
abundance of hot gas. Henceforth, we shall 
neglect the last term in equation (4) 
(imposing $\dot{M}_{\rm s} \equiv 0$), i.e., we shall 
assume that the amount of gas deposited in 
a dusty disk or sequestered into new stars (and then 
new compact remnants) is negligible compared 
to the accretion and outflow components.

In the simplest scenario, with constant accretion 
rate (i.e., $a$ and $b$ constant) 
and constant wind losses, the total gas 
mass inside the nuclear region would 
increase linearly with time unless the outflows 
are fine-tuned to balance the gas sources:
\begin{equation}
\dot{M}_{\rm w} = (1-a) \dot{M}_{\rm B} +  (1-b) \dot{M}_{\ast}.
\end{equation}
To remove the excess gas from the SMBH sphere 
of influence, it has to be imparted a velocity 
\begin{eqnarray}
v_{\rm w} &\ge& v_{\rm esc} \approx 
(2GM_{\rm BH}/r)^{1/2}\nonumber\\ 
&\approx& 320 
(r/r_{\rm acc})^{-1/2}(kT/0.5 {\rm ~keV})^{1/2} {\rm ~km~s}^{-1},
\end{eqnarray}
spending a power 
\begin{eqnarray}
P_{\rm w} = (1/2) \dot{M}_{\rm w} v_{\rm w}^2 &\approx & 
7.9 \times 10^{37} 
\left(\frac{v_{\rm w}}{500{\rm~km~s}^{-1}}\right)^2 \times \nonumber\\
&&
\left(\frac{\dot{M}_{\rm w}}{10^{-3} M_{\odot} {\rm ~yr}^{-1}}\right) 
	\ {\rm erg~s}^{-1}.
\end{eqnarray}

For our target galaxies, this is perfectly compatible 
with the energy budget in the system. For example, 
one source of energy readily available comes from 
Type-Ia SNe, from the same stellar population 
responsible for the gas injection $\dot{M}_{\ast}$.
The total SN heating can be estimated as 
\begin{equation}
L_{\rm SN}(t) \approx 7.1 \times 10^{30} \theta_{\rm SN} 
	\left(\frac{L_{B}}{L_{\odot, B}}\right) 
	\left(\frac{t}{15 {\rm ~Gyr}}\right)^{-s} \ {\rm erg~s}^{-1}
\end{equation}
(Ciotti et al.~1991), where $\theta_{\rm SN} \la 1$ 
and $s \ga 1.3$. For our target galaxies, typical numbers 
are SN rates $\sim 10^{-5}$ yr$^{-1}$ inside $r_{\rm acc}$, 
implying heating rates of a few $10^{38}$ erg s$^{-1}$.
From equations~(2), (8) and (9), we obtain that
only a fraction $\kappa \approx 0.17 \times 
[v_{\rm w}/(500 {\rm~km~s}^{-1})]^2$ 
of this power is needed to remove 
$\dot{M}_{\rm w} \approx \dot{M}_{\ast}$ from the SMBH 
sphere of influence.

Although consistent with the energy available, 
this scenario requires an uncomfortable degree 
of fine-tuning between injection, outflows and accretion.
To make it more plausible, we need to make both 
the accretion and outflow rates self-regulated, 
in order to achieve a quasi-steady state.
Firstly, to prevent the gas from building up 
indefinitely, we assume that 
the accretion rate is a function of the gas density 
inside or at the surface of the sphere of influence, 
and therefore, as a first approximation, 
to the total gas mass inside $r_{\rm acc}$.
We already know that this is true for Bondi accretion 
($\dot{M} \sim \dot{M}_{\rm B} \sim n_e$, Paper I; Bondi 1952).
If this holds for all gas components, we can write 
$\dot{M}(t) \equiv b' M_{\rm a}(t)$ in Equation (4).
The parameter $(1/b')$ is the timescale on which 
the sphere of influence would be drained if all 
gas sources and outflows are turned off. This is of order 
of the free-fall timescale inside the region,  
$t_{\rm ff} \sim  r_{\rm acc}^{3/2} M_{\rm BH}^{-1/2}$ 
(or a factor of 10 larger, for ADAF-like inflows; 
Narayan \& Yi 1994), and is essentially 
the same timescale used to estimate the Bondi 
accretion rate. 
Recalling the physical scaling of $r_{\rm acc}$ 
(eq.~[5] in Paper I), we can write 
\begin{equation}
b' \approx 2.5 \beta \times 10^{-5} 
\left(\frac{10^8 M_{\odot}}{M_{\rm BH}}\right)
\left(\frac{kT}{0.5 {\rm ~keV}}\right)^{3/2} \ \ {\rm yr}^{-1},
\end{equation}
where $0.1 \la \beta \la 1$ depending on the details 
of the accretion flow.
Secondly, to explain how the wind can be fine-tuned 
to remove $\sim 90$--$99\%$ of the injected gas, 
we assume that the net outflow 
rate is naturally tied to the accretion rate, 
as described in the next section.

\subsection{Black hole feedback: outflows and jets}

Feedback from the SMBH, either via deposition 
of mechanical energy or through heating of the ISM, 
is a possible way to stop inflows 
and/or increase the rate of gas outflows  
from the nuclear region 
(Ostriker \& Ciotti 2005; Pellegrini 2005; Omma et al 2004; 
Ciotti \& Ostriker 2001; Binney \& Tabor 1995).
We can model this scenario by assuming 
that a fraction $\kappa' (\eta'\dot{M}c^2) 
= \kappa' \eta' b'M_{\rm a}(t)c^2$ of the 
accretion power is recycled to eject gas 
from the nuclear region; here $\eta'$ is 
the total efficiency (radiative plus mechanical) 
because both kinds of BH feedback may 
contribute to the driving of an outflow.
Similarly to our previous argument, 
we propose that this feedback power will 
impart a (mass-averaged) velocity 
$v_{\rm w} \ge v_{\rm esc}$ 
to the outflowing gas.  Hence, we require that 
\begin{equation}
(1/2)\dot{M}_{\rm w}v_{\rm w}^2 = \kappa' \eta' b'M_{\rm a}(t)c^2,
\end{equation}
from which we obtain
\begin{eqnarray}
\dot{M}_{\rm w}(t) &\approx& 7200 \left(\frac{\kappa'}{0.1}\right) 
	\left(\frac{\eta'}{0.1}\right) 
	\left(\frac{v_{\rm w}}{500 {\rm ~km~s}^{-1}}\right)^{-2}
	b'M_{\rm a}(t)\nonumber\\ 
	&\equiv& p b'M_{\rm a}(t),
\end{eqnarray}
where we have taken $\kappa' \sim 10\%$ 
as a characteristic fraction of energy feedback, 
and $p$ is a numerical parameter.
By inserting $\dot{M}_{\rm w}(t)$ from equation~(12) into equation~(4), 
and solving, we obtain the asymptotic quasi-steady-state values 
of mass, average electron density, and accretion rate 
inside the sphere of influence:
\begin{eqnarray}
M_{\rm a}(t=\infty) &\approx& \frac{40}{\beta (p+1)} 
\left(\frac{M_{\rm BH}}{10^8 M_{\odot}}\right)
	\left(\frac{0.5 {\rm ~keV}}{kT}\right)^{3/2}\nonumber\\
	&&\times
\left(\frac{\dot{M}_{\rm B} + \dot{M}_{\ast}}{10^{-3} 
	M_{\odot} {\rm ~yr}^{-1}}\right) \ \ M_{\odot},\\
n_{\rm e}(t=\infty) &\approx& \frac{0.51}{\beta (p+1)} 
\left(\frac{10^8 M_{\odot}}{M_{\rm BH}}\right)^{2}
	\left(\frac{kT}{0.5 {\rm ~keV}}\right)^{3/2}\nonumber\\
	&&\times
\left(\frac{\dot{M}_{\rm B} + \dot{M}_{\ast}}{10^{-3} 
	M_{\odot} {\rm ~yr}^{-1}}\right) \ \ {\rm cm}^{-3},\\
\dot{M}(t=\infty) &=& (\dot{M}_{\rm B} + \dot{M}_{\ast})/(p+1).
\end{eqnarray}
In the basic ADAF scenario, $p \approx 1/\alpha \approx 10$;
$p \sim 10$--$10^3$ in CDAF or ADIOS scenarios, which 
predict a lower mass rate reaching the central BH.
These values are consistent with the observations 
(Col.10 in Table 3).

In terms of the feedback coupling constant $k'$, 
we obtain:
\begin{equation}
\left(\frac{\kappa'}{0.1}\right) \approx  
	\left(\frac{p}{7200}\right)
	\left(\frac{0.1}{\eta'}\right)
	\left(\frac{v_{\rm w}}{500 {\rm ~km~s}^{-1}}\right)^{2}.	
\end{equation}
If this feedback coupling 
can be achieved (i.e., if $\kappa' \la 1$), 
the SMBH feedback suffices to unbind and remove 
the excess gas produced by stellar winds in the nuclear region.
For a purely advective inflow, 
$(0.1p/\eta') \approx (0.1p/\eta) \sim \alpha/\dot{m} \sim 10^4$, 
hence $0.1 \la \kappa' \la 1$.
In other types of radiatively-inefficient flows, 
including ADIOS, CDAF and jet systems, 
where a smaller fraction of the accretion power is advected, 
more accretion power is available to drive an outflow, 
so that $\kappa' \ll 0.1$.

In our phenomenological model, we require  
that slow ($v_{\rm w} \sim v_{\rm esc}$) 
outflows carry out a fraction $p/(p+1)$  
of the available gas mass (not considering 
a possible mass component deposited in a dusty disk  
or new stars), while only a fraction 
$1/(p+1)$ reaches the SMBH. This ratio is somewhat 
reversed if we look at the power budget. The same 
outflows consume only a fraction $\kappa'$ 
of the available total (radiative plus mechanical) 
accretion power, with $\kappa' \ll 1$ for many 
varieties of radiatively-inefficient 
solutions. The remaining fraction $(1-\kappa' - f_{\rm r})$ 
of the accretion power $P_{\rm acc} = \eta'\dot{M}c^2$  
is still available, for example, 
for launching a fast jet. 

The kinetic energy carried 
by a relativistic jet is 
$P_{\rm J} = \gamma_{\rm J} \dot{M}_{\rm J} c^2$.
A bulk Lorentz factor $3 \la \gamma_{\rm J} \la 10$ 
has been observed in quasars and AGN (Falcke \& Biermann 1995; 
Falcke, Malkan \& Biermann 1995; Dopita 1997).
A similar range of Lorentz factors is observed 
in stellar-mass BHs in the ``very-high state'' 
(Fender et al.~2004). Instead, the range of Lorentz 
factors for steady jets in the ``low/hard state'' 
(a case more relevant to our sample of 
quiescent or low-luminosity SMBHs) 
is still hotly debated. Heinz \& Merloni (2004) 
inferred a lower limit $\gamma_{\rm J} \ga 5$ 
if the X-ray emission is unbeamed;
on the other hand, Fender et al.~(2004) 
argued that jets in the low/hard state are only 
mildly relativistic, with $v_{\rm J} \la 0.6 c$, 
$\gamma_{\rm J} \la 1.4$.

If the power budget of X-ray-faint SMBHs 
is dominated by a relativisitc jet, 
we expect the mass carried by the jet 
to be $\approx (\eta'/\gamma_{\rm J}) \dot{M}$, 
i.e., only $\sim 1\%$ of the mass that goes 
into the BH, and $\la 0.1\%$ of the mass carried away 
by the slow winds. On the contrary, 
if the jet is slower, for example with $v_{\rm J} \approx 0.5 c$, 
its kinetic energy is $\approx (1/8) \dot{M}_{\rm J} c^2$, 
and therefore it can carry a mass $\dot{M}_{\rm J} \sim \dot{M}$, 
i.e., it can also contribute significantly 
to the mass outflow. Measuring the radio core luminosity 
of our target galaxies would provide 
an additional constraint to the model.

\subsection{Correlation between X-ray luminosity and accretion rate}

In the framework of our phenomenological model, 
we can now also re-interpret the source distribution 
in the X-ray-luminosity versus Bondi-accretion-rate plane 
(cf.~Figure 14 in Paper I). We speculate that the apparent lack 
of any correlations, noted by Pellegrini (2005), 
and that we confirmed with our additional data, 
is a consequence of the fact that the X-ray luminosity 
in the systems is a function of three physical parameters, 
rather than just one or two, as initially hypothesized.
The three parameters are: the hot-ISM inflow rate 
into the SMBH sphere of influence 
(approximated by the Bondi rate, $\dot{M}_{\rm B}$); 
the stellar mass losses inside the SMBH sphere of influence 
($\dot{M}_{\ast}$); and the fraction $1/(p+1)$ 
of the total gas available that actually accretes 
onto the SMBH. 
 
We have replotted the same data points in Figure 5, 
and this time we have considered the dependence 
on all three parameters. We speculate that 
data points on the left-hand side of the plot 
are characterized by $\dot{M}_{\ast} 
\gg \dot{M}_{\rm B}$, and hence $\dot{M} \ga \dot{M}_{\rm B}$ 
and, adopting the self-similar ADAF efficiency (Section 3.2), 
\begin{equation}
\frac{L_{\rm X}}{L_{\rm Edd}} \approx 
\left[\frac{\dot{M}_{\ast}}{\alpha (p+1) \dot{M}_{\rm Edd}}\right]^2 
\sim 10^{-6} {\rm -} 10^{-8}, 
\end{equation}
independent of $\dot{M}_{\rm B}$. 
Data points on the right-hand side of the diagram have  
$\dot{M}_{\rm B} \gg \dot{M}_{\ast}$, 
and hence $\dot{M}/\dot{M}_{\rm B} \approx 1/(p+1) \ll 1$
and 
\begin{eqnarray}
\frac{L_{\rm X}}{L_{\rm Edd}} &\approx &
\left[\frac{\dot{M}_{\rm B}}
{\alpha (p+1) \dot{M}_{\rm Edd}}\right]^2,\nonumber\\
\log \left(\frac{L_{\rm X}}{L_{\rm Edd}}\right)   &\approx &
	2 \log \left(\frac{\dot{M}_{\rm B}}{\dot{M}_{\rm Edd}}\right) 
	- 2 \log [\alpha (p+1)],	
\end{eqnarray}
with $0 \la \log [\alpha (p+1)] \la 1$.
We have shown the expected location of the 
nuclear X-ray sources for two choices of stellar mass 
losses ($\dot{M}_{\ast}/\dot{M}_{\rm Edd} = 10^{-4}$: 
dashed lines;  
and $\dot{M}_{\ast}/\dot{M}_{\rm Edd} = 10^{-3}$: solid lines) 
and for three possible values of 
the accretion fraction: $1\%$ (indicative of  
ADIOS or CDAF solutions), $10\%$ (ADAF), and $50\%$.
In particular, at least five of our six target galaxies 
have $\dot{M}_{\ast}/\dot{M}_{\rm Edd} \sim$ a few $10^{-4}$ 
and are consistent with an accretion fraction $1/(p+1) \sim 10\%$ 
(Table 3).


\begin{figure*}[t]
\includegraphics[angle=270,scale=0.65]{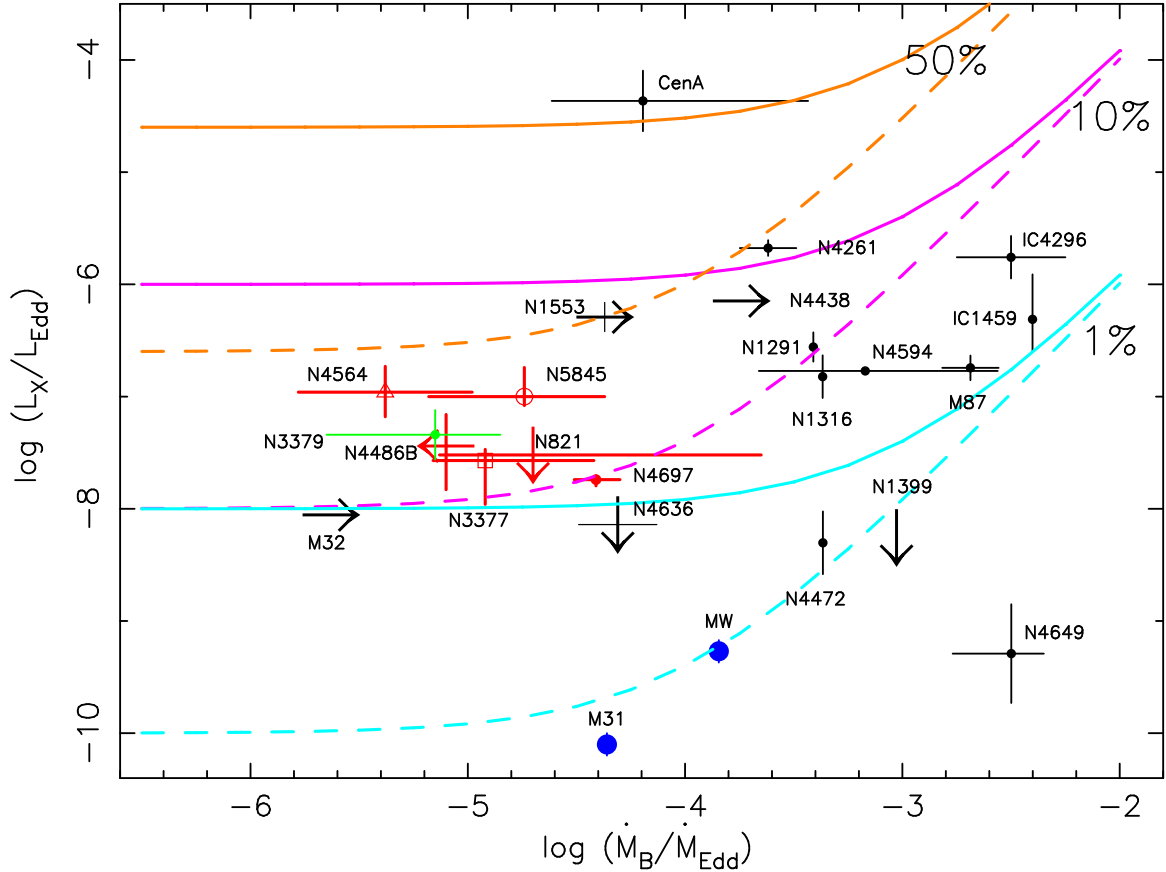}
\caption{Normalized X-ray luminosity 
$L_{\rm X}/L_{\rm Edd}$ as a function of normalized 
Bondi accretion rate 
$\dot{m}_{\rm B} \equiv \dot{M}_{\rm B}/\dot{M}_{\rm Edd}$ 
(cf.~Figure 14 in Paper I).
Here we have taken into account the presence 
of stellar mass losses (winds) as an additional source of gas. 
We have plotted the expected luminosities for two 
``plausible'' choices of normalized 
stellar mass losses inside the accretion radii: 
$\dot{M}_{\ast}/\dot{M}_{\rm Edd} = 10^{-4}$ (dashed lines)
and $\dot{M}_{\ast}/\dot{M}_{\rm Edd} = 10^{-3}$ (solid lines); 
this is a typical range of values inferred  
for galaxies for which the optical brightness 
profiles of the nuclear regions are available  
(e.g., Table 3; and Fig.~1 in Gebhardt et al.~2003).
For each stellar mass loss rate, we have plotted the luminosity 
expected for three different values of $1/(1+p)$, which 
parameterizes the total fraction of gas available that 
is actually accreted onto the SMBH ($1\%$, $10\%$, or $50\%$).
We have compared these curves with the observed 
values: $L_{\rm X}$ and $\dot{M}_{\rm B}$ 
are obtained from X-ray observations. 
Data points in red (in the online version) 
are the six new galaxies 
of our study; data points in black are the galaxies 
in Pellegrini's (2005) sample; the data point 
in green is from David et al.~(2005); the large,
filled (blue) circles are Sag A$^{\ast}$ 
(Baganoff et al.~2003) and M\,31$^{\ast}$ 
(Garcia et al.~2005).
Assuming a characteristic gas injection rate 
$\dot{M}_{\ast}/\dot{M}_{\rm Edd} \sim$ a few $10^{-4}$, 
we conclude that most SMBHs are 
consistent with an accretion fraction $\sim 1$--$10\%$. 
However, even lower accretion fractions 
are inferred for NGC\,4649; conversely, NGC\,5128 (Cen A) 
is consistent with an accretion fraction $\approx 50\%$. 
Another two galaxies (NGC\,4486B and NGC\,4261) 
are consistent with accretion fractions $\sim 20$--$50\%$.
\label{fig18}}
\end{figure*}


\subsection{Intermittent accretion and state transitions}

So far, we have assumed that the system is in a steady state.
It is possible to generalize our simple model  
in such a way that the system goes through a cycle 
of low and high luminosity and low and high 
gas density (intermittent accretion, see Section 5.4 in Paper I,  
and Pellegrini 2005), 
especially if the outflows are powered by SMBH feedback. 
A general condition is that matter 
would alternatively be accumulated 
inside $r_{\rm acc}$ (a phase in which $\dot{M} + \dot{M}_{\rm w} 
\ll \dot{M}_{\rm B} + \dot{M}_{\ast}$) 
and depleted ($\dot{M} + \dot{M}_{\rm w} 
\gg \dot{M}_{\rm B} + \dot{M}_{\ast}$).
This can be obtained for example by assuming 
that the feedback coupling strength $\kappa'$ 
depends on the luminosity or the mass accretion rate, 
being $\ll 0.01$ below a certain accretion rate 
(thus, unable to drive a wind) and $> 0.1$ above that threshold, 
with the addition of hysteresis or of a time lag 
with respect to the SMBH activity. (For example, 
the time lag could be due to the cooling timescale 
of the ISM heated/shocked by the SMBH activity.)

A different way of explaining intermittent 
episodes of nuclear activity 
is to invoke changes in the relative fraction 
of accretion power carried outwards as kinetic energy
by a relativistic jet (Fender, Gallo, \& Jonker 2003). 
We recall 
that, while most of the mass may be removed by a slow wind, 
most of the accretion power is either advected into the SMBH 
or carried outwards by a fast jet. Relativistic electrons 
in a jet may be responsible for inverse-Compton scattering  
of ambient optical/UV photons; the interaction of a jet 
with the surrounding ISM, the accretion inflow, 
or the slow wind itself will produce strong shocks, 
and associated synchrotron emission. Hence, 
changes in the relative fraction of power in the 
jet will produce noticeable effects in both the 
radio and X-ray emission.

Another mechanism for intermittent accretion is a transition 
in the physical properties of the accretion flow, 
above a certain density threshold. In fact, it has been 
suggested that changes in the jet power and in the inflow 
structure are physically correlated (Fender et al.~2004; 
Merloni et al.~2003). 
Siemiginowska, Czerny, \& Kostyunin (1996) modelled the effect 
of thermal-viscous instabilities in an accretion disk, 
which can switch between a cold state (lower viscosity 
and $\dot{M}$) and a hot, fully ionized state 
(higher viscosity and $\dot{M}$), and suggested 
that X-ray faint SMBHs may be in the low state.

Another possible transition is the one between 
a radiatively inefficient and a radiatively efficient 
solution (standard disk, Shakura \& Sunyaev 1973). This is 
expected to occur at mass accretion 
rates $\dot{m} \sim 10^{-2}$, at which point 
the ADAF efficiency matches the standard efficiency.
This corresponds to a mass injection rate
\begin{equation}
\left(\dot{M}_{\rm B} +\dot{M}_{\ast}\right) \sim 2.3 \times 10^{-2} (p+1)
	\left(\frac{M_{\rm BH}}{10^8 M_{\odot}}\right) \ \ 
	M_{\odot} {\rm ~yr}^{-1}.
\end{equation}
If Bondi accretion of the hot ISM is the dominant component, 
this implies number densities at $r = r_{\rm acc}$ 
\begin{equation}
n_e(r_{\rm acc}) \ga 14.6 (p+1)
	\left(\frac{kT}{0.5 {\rm ~keV}}\right)^{3/2}
	\left(\frac{M_{\rm BH}}{10^8 M_{\odot}}\right)^{-1} \ \ 
	{\rm cm}^{-3},
\end{equation}
corresponding to hot gas masses $\ga 10^3 M_{\odot}$ 
inside the sphere of influence of the BH, 
for uniformely distributed gas.

Quiescent SMBHs may in principle reach this threshold 
at some stage, if the gas inflow rate is as high 
as in equation~(19); or even for a lower injection rate, 
if accretion plus outflows plus any other forms 
of mass sequestration cannot balance the gas inflow, 
so that gas keeps building up inside the sphere of influence. 
At that point, radiative cooling inside $r_{\rm acc}$ 
becomes efficient, and the gas 
collapses into a standard disk; hence, we may expect  
the accretion rate to increase or jump to a value   
$\dot{M} \gg (\dot{M}_{\ast} - \dot{M}_{\rm w})$, 
until the disk is drained and the nuclear region 
is depleted of its gas (not only through disk accretion, 
but also through more efficient radiatively-driven 
outflows). 
During this outburst, the nucleus 
will appear as a bright AGN, with 
a luminosity $L_{\rm bol} \sim \dot{m} L_{\rm Edd} 
\sim 10^{44}$ erg s$^{-1}$. After the disk is emptied, 
accretion resumes in a low, radiatively-inefficient 
regime, and the gas mass keeps building up again 
until the following outburst.
This scenario may be applicable to the nuclei 
of moderately bright systems ($L_{\rm X} 
\sim 10^{41}$--$10^{43}$ erg s$^{-1}$) but 
it is almost certainly not relevant to our sample 
of faint elliptical galaxies, and probably not 
for any galaxy of Pellegrini's (2005) sample either.
Current gas densities and accretion rates 
in our target nuclei are very far (three orders of magnitude) 
from the transition threshold. 

It is possible that, in some cases, the accretion 
is out of equilibrium and gas may be starting 
to build up again after an outburst 
has recently emptied the nuclear region. 
If so, applying our phenomenological model 
(Section 4.3), 
we estimate that the mass accretion rate will 
increase, reaching an asymptotic value 
within $\la 10^6$ yr.
However, the observed X-ray luminosities for our target nuclei  
point to accretion rates $\sim 10^{-5} M_{\odot}$ yr$^{-1}$, 
one or two orders of magnitude lower than the injection rate.
It is statistically unlikely that we are observing 
all our targets in the first $\sim 10^3$--$10^4$ yr after each 
outburst; we would, instead, expect to find most 
of the galaxies close to their efficient-accretion threshold or 
to their asymptotic maximum accretion rate, at any given time.
We are aware that ours is a biased sample, 
because we have chosen some of the X-ray faintest 
among the nearby galaxies with an SMBH mass determination.
But we do have estimates of the SMBH 
luminosity and accretion rate for about $1/3$ 
of the $\approx 50$ galaxies with a known SMBH mass: 
in almost all of them, the accretion rate 
is orders of magnitude lower than the 
efficient-accretion threshold. 
We conclude that they cannot all be out of equilibrium 
or in the initial phase of a transient cycle. 

Even if our quiescent target nuclei 
could accrete all the inflowing hot and warm gas 
(Table 3) without any outflows, the mass accretion rate
would still be at least an order of magnitude too low 
to trigger a transition from radiatively inefficient 
to radiatively efficient flows.
Additional sources of gas, such as a more powerful 
cooling flow or satellite accretion from the outer galactic 
regions, or a much younger stellar population 
in the inner regions, would be required 
for the system to reach the efficient-accretion threshhold.
For example, a stellar population with characteristic 
age $\la 500$ Myr (equation 2, see also Ciotti et al.\ 1991) 
would provide a mass-loss rate more than an order 
of magnitude higher, that is probably 
high enough to permit at least transient radiatively-efficient 
accretion.

\section{Summary and conclusions}

We have studied a sample of X-ray faint early-type galaxies 
with {\it Chandra} (see Paper I), {\it HST} and ground-based 
optical data, with the objective of clarifying 
the relation between X-ray luminosity of their SMBHs, 
total accretion rate, radiative efficiency, 
and classical Bondi rate of capture 
of the hot ISM by the SMBH sphere of influence.
We have focused on our new results for a sample of 
six galaxies with kinematic SMBH mass determinations; 
in addition, we have used the data available 
in the literature for a larger sample of galaxies  
(Pellegrini et al.~2005; Garcia et al.~2005; 
David et al.~2005).

The low SMBH X-ray luminosities rule out radiatively 
efficient, standard accretion.
We noted in Paper I that radiatively-inefficient 
accretion models provided better estimates, but 
the X-ray data did not show a clear relation between 
Bondi inflow rates and nuclear luminosities.
In a few cases, the accretion rates required 
to match the X-ray luminosities  
are $\ll 10\%$ of the Bondi rate; this supports 
radiatively-inefficient flows with 
convection and/or outflows, rather than 
simple advection. In other cases, 
the required rates are $\sim 10\%$ of the Bondi rate, 
consistent with basic ADAF models. In a few more cases, 
and in particular for at least five of our six new targets, 
the accretion rates must be $\ga$ Bondi rate; this is 
difficult to reconcile with the prediction 
of radiatively inefficient models.

In this work, we explained this discrepancy 
by using a new empirical method to estimate 
the gas injection rate 
into the SMBH sphere of influence. We suggested  
that it can be expressed as the sum of the hot-ISM 
inflow rate (estimated from {\it Chandra} 
observations of the diffuse hot gas, 
with spatial resolutions $\ga 100$ pc; see Paper I), 
plus additional contributions 
from the stellar population 
inside the sphere of influence
(stellar winds and Type-Ia SNe). The stellar 
contribution can be estimated by deprojecting 
the optical brightness profiles 
to obtain the volumetric luminosity densities, and 
applying standard relations between optical luminosity, 
stellar densities and ages, and mass loss rates.
We found typical stellar mass loss rates 
$\sim 10^{-4}$--$10^{-3} M_{\odot}$ yr$^{-1}$; 
on the other hand, the hot gas content varies greatly, 
leading to X-ray-estimated Bondi rates 
from $\la 10^{-5}$ to $\sim 10^{-2} M_{\odot}$ yr$^{-1}$ 
over the full sample of galaxies.

We have used these two parameters ($\dot{M}_{\rm B}$ 
and $\dot{M}_{\ast}$) to model the total mass 
injection rate. Only an {\it a priori} unknown 
fraction of this gas reaches the SMBH 
(the rest being re-ejected, stored, or turned 
into new stars), which adds another parameter 
to the model. And only an {\it a priori} unknown 
fraction of the accretion power is released 
as X-ray flux (the rest being advected or carried out 
as mechanical luminosity, in a radio jet or a wind).
Various accretion flow solutions (standard disk, ADAF, etc.) 
have different predictions for the fraction of gas accreted 
by the BH, and for its radiative efficiency.
Assuming the ADAF radiative efficiency 
$\eta \approx 10 \dot{M}/\dot{M}_{\rm Edd}$, the observed  
X-ray luminosities imply that, 
for most galaxies, only $\sim 1$--$10\%$ 
of the inflowing gas is accreting onto their SMBHs.
We suggested that the intrinsic scatter 
and/or observational uncertainty in $\dot{M}_{\ast}$ 
and in the accretion fraction is the main 
reason for the lack of correlation 
between the Bondi rate 
and the X-ray luminosity of the SMBH.
 
Measuring the core radio luminosity of 
an SMBH offers a possible way 
of determining what fraction 
of the accretion power is advected 
and what is instead carried outwards 
as mechanical luminosity by a compact jet, 
or used to power convective flows and outflows
(Merloni et al.\ 2003).  
We only have upper limits to the core radio emission 
of our target galaxies. Most of them lie 
somewhat below the Merloni-Heinz-DiMatteo 
fundamental-plane correlation, suggesting that 
advection, slow outflows, and convection 
may be more important than relativistic jets; 
however, deeper radio observations 
would be required to provide firm conclusions.


Based on our X-ray and optical results, we have discussed 
the conditions 
for mass equilibrium inside the SMBH sphere of influence 
and the fate of the gas that does not sink into the SMBH.
It is possible that 
in some cases (most notably in NGC\,5845), 
part of the excess gas may cool down, settle 
into a dusty/stellar disk, and eventually 
form new stars, even inside the Bondi accretion 
radius. In fact, a global understanding 
of SMBH physics (both active and quiescent) 
requires simultaneous modelling of all three 
phases (accretion, ejection and star formation). 
However, for this work, we used the simplifying 
assumption that star formation is negligible 
and slow, massive outflows remove $\sim 90$--$99\%$ 
of the gas from the nuclear region.
 
Type-Ia SNe could in principle provide 
enough power to heat the gas and remove it 
from the central region. However, reaching 
a steady state would require an implausible degree 
of fine-tuning between injection, accretion, 
and outflow rates. A more likely scenario is that 
both the accretion rate and the outflow rate 
are self-regulating: this happens, for example,  
if the accretion rate is proportional to the gas 
density or total mass inside the sphere of influence 
of the SMBH, and the power carried by the outflow 
is proportional to the accretion power. In this case, 
the system can reach an asymptotic equilibrium; 
this mechanism relies on the idea of SMBH feedback.
We have estimated what fraction of the accretion power 
has to be used to drive away the excess gas. Pure ADAF solutions 
require a highly efficient feedback coupling, 
essentially because most of the accretion power 
is lost into the BH. Other radiatively inefficient 
solutions only require that $\la 1\%$ 
of the available accretion power be used for 
the feedback. Hence, a fast jet 
may still carry outwards $\ga 99\%$ of the accretion 
power. If the jet is fully relativistic 
($\gamma_{\rm J} \ga 5$: Heinz \& Merloni 2004), 
it can carry outwards $\sim 0.1\%$ of the inflowing mass, 
with the rest being either accreted or removed 
by feedback-driven, slow outflows. 
If the jet is only midly relativistic 
($v_{\rm w} \la 0.5 c$: Fender et al.~2004), 
it can carry away an amount of mass 
comparable to what sinks into the SMB, 
$\sim 10\%$ of the inflowing gas, with 
the difference being removed by 
a slower outflow component. 
As an aside, we note that a typical outflow rate of 
$\sim 10^{-3} M_{\odot}$ yr$^{-1}$ 
from inside the sphere of influence would, 
over a Hubble time, inject 
a gas mass of $\sim 10^7 M_{\odot}$ 
into the galaxy: this is 
negligible compared to other sources of gas.

Alternatively, one can propose scenarios 
in which the system may never reach 
a steady state: the coupling of the central engine 
with the surrounding ISM may switch between 
an efficient and an inefficient regime, perhaps  
with a time lag or hysteresis, such that 
gas alternatively builds up or gets drained 
from the SMBH sphere of influence.
For example, when active, the SMBH could be 
accreting from the gas accumulated in the inner 
few pc, while at the same time suppressing further 
inflows from larger radii. Observational evidence 
for feedback-driven intermittent accretion 
in NGC\,821 was discussed in Fabbiano et al.~(2004);
see also Pellegrini (2005).

Intermittent accretion may also be triggered 
by transitions in the nature of the accreting 
inflow, regardless of SMBH feedback. For example, 
it could be due to a switch from 
a radiatively inefficient to a radiatively 
efficient solution, with the collapse 
of the gas into an optically thick 
accretion disk; or to a thermal-viscous 
disk instability. Both mechanisms require 
accretion rates $\ga 10^{-2} \dot{M}_{\rm Edd}$, 
two orders of magnitude higher 
than what is currently observed 
in our quiescent target galaxies. Instead, they may apply 
to younger elliptical galaxies ($\la 500$ Myr), 
or to galaxies where most of the accreting gas 
is provided by stronger cooling flows from larger radii, 
or that have undergone recent satellite accretions.

In this paper, we have only considered the energy 
and mass balance inside the sphere of influence 
(and neglecting star formation).
A more detailed analysis of the conditions for 
SMBH feedback must include a study 
of the timescale on which feedback mechanisms 
operate (e.g., Ostriker \& Ciotti 2005; 
Sazonov et al.~2005). For example, one has 
to compare the timescale for Compton heating 
from the central X-ray source with 
the accretion timescale. On the other hand, 
if the SMBH has a strong UV flux component, 
line-driven winds would be more effective  
to push out the gas originating 
from stellar mass losses (warm phase), 
before it has time to join the hot phase. 
Shock heating from a jet would also provide 
effective feedback. 

Another related complication we have neglected 
here is that SMBH feedback, in addition 
to removing gas, can also heat up the gas 
that remains within the Bondi sphere, 
thus making its accretion more difficult 
(eq.[6] in Paper I). 
In this paper we have simply assumed that 
the diffuse hot gas is approximately isothermal 
over the whole galaxy (see Paper I). 
If the temperature has a spike inside the Bondi 
radius (not resolved by {\it Chandra}) 
due to SMBH feedback, part of the point-like 
hard X-ray emission may in fact be due to
the gas itself. In any case, it remains 
true that, in a steady state, the eventual fate 
of this heated gas is to be either accreted 
or ejected. If it just kept building up, 
its increased density would make cooling
more and more efficient (cooling rate $\sim n_e^2$).
Finally, if nuclear 
star formation (to be studied with 
{\it Spitzer}) is significant, that may also 
be regulated by SMBH feeedback. 
A study of these issues is left 
to futher work.

As a serendipitous result of our study, we have estimated 
an alternative SMBH mass for the double-nucleus, 
compact elliptical galaxy NGC\,4486B. Using the correlations 
between SMBH mass and S\'{e}rsic index $n$ of the optical brightness 
profiles, and with the stellar velocity dispersion, we have argued 
that its BH mass may be a factor of 10 lower 
than the value obtained from two-integral 
models of its stellar kinematics.

\acknowledgments

We are grateful to Nicola Caon who provided us with the
calibration for several of the optical light-profiles.
We also thank Alice Quillen for supplying us
with the (inner) near-IR light-profile, and Armin Rest
for supplying the (inner) light-profile for NGC\,3377.
We thank Andreas Zezas for his suggestions 
on the X-ray data analysis and interpretation, 
and Hermine Landt for discussions on the mass balance. 
Thanks also to the anonymous referee 
for his/her useful comments and suggestions.
This work was supported by the NASA Chandra Guest Observer 
grant GO3-4133X and by the NASA contract NAS8-39073 (CXC).
RS acknowledges partial support from a Marie Curie fellowship.

\end{document}